\PassOptionsToPackage{table}{xcolor}
\documentclass[fleqn,10pt]{wlscirep}
\usepackage[utf8]{inputenc}
\usepackage[T1]{fontenc}

\usepackage{graphicx}
\usepackage{subcaption} 
\usepackage[table]{xcolor} 
\usepackage{multirow}
\usepackage{lettrine}
\usepackage{enumitem}

\usepackage{float}
\usepackage{caption}
\usepackage{amsmath}
\usepackage{hyperref}
\usepackage{lettrine}

\usepackage{fancyhdr}
\usepackage{refcount}

\usepackage{algorithm}
\usepackage{algorithmic}
\usepackage{amsmath}
\usepackage{amssymb}
\usepackage{todonotes}
\usepackage{rotating}
\usepackage{lineno}

\usepackage{pdfpages}
\usepackage{booktabs}
\usepackage{array}
\usepackage[table]{xcolor}
\usepackage{tikz}
\usepackage{threeparttable}

\usepackage{placeins}

\setlist[itemize]{noitemsep}

\definecolor{mygray}{gray}{0.95}
\definecolor{myblue}{rgb}{0, 0, 0.99}

\definecolor{headgray}{gray}{0.90}
\definecolor{rowgray}{gray}{0.95}
\definecolor{dotred}{RGB}{222,85,82}

\newcommand{\rdot}{\tikz[baseline=-0.6ex]\fill[dotred] (0,0) circle (1.2mm);}
\newcommand{\rodot}{\tikz[baseline=-0.6ex]\draw[dotred,line width=0.5pt] (0,0) circle (1.2mm);}
\usepackage{hyperref}
\usepackage{shellesc}
\newif\ifcountwords

\countwordstrue

\title{Benchmarking bandgap prediction in semiconductors under experimental and realistic evaluation settings}

\author[1,2]{Haolin Wang}
\author[1,2]{Xianyuan Liu}
\author[3]{Anna Jungbluth}
\author[4]{Alexandra J. Ramadan}
\author[5]{Robert D. J. Oliver}
\author[1,2,*]{Haiping Lu}
\affil[1]{Centre for Machine Intelligence, University of Sheffield, Sheffield, S1 3JD, UK}
\affil[2]{School of Computer Science, University of Sheffield, Sheffield, S1 4DP, UK}
\affil[3]{Climate Office, European Space Agency, Harwell, OX11 0FD, UK}
\affil[4]{School of Mathematical and Physical Sciences, University of Sheffield, Sheffield, S3 7RH, UK}
\affil[5]{School of Chemical, Materials and Biological Engineering, University of Sheffield, Sheffield, S1 3JD, UK}

\affil[*]{Corresponding author: Haiping Lu (h.lu@sheffield.ac.uk)}

\begin{abstract}
Accurate bandgap prediction is crucial for semiconductor applications, yet machine learning models trained on computational data often struggle to generalize to experimental bandgap measurements. Challenges related to data fidelity, domain generalization, and model interpretability remain insufficiently addressed in existing evaluation frameworks. To bridge this gap, we introduce RealMat-BaG, a benchmark for assessing model reliability under experimentally relevant conditions. We curate an open-access dataset of experimental bandgaps with aligned crystal structures and compare graph neural networks as well as classical machine learning baselines. Our framework evaluates performance across statistical and domain-based splits, examines transfer from DFT-computed to experimental bandgaps, and analyzes interpretability at both elemental-property and structural levels. Our results reveal the fundamental generalization limitations of current bandgap prediction models and establish a benchmark aligned with experimental measurements for developing more reliable learning strategies for materials discovery.

\end{abstract}

\begin{document}

\flushbottom
\maketitle

\thispagestyle{empty}

\section*{Introduction}

The bandgap is a fundamental property of semiconductors. Determining the bandgap is crucial because it defines a material’s electronic and optical behavior and serves as a primary filter for assessing suitability in devices such as transistors, LEDs, and solar cells \cite{Yoder1996WideSemiconductor, Ueno2004Field, Lisensky1992Periodic, Licht2001Multiple}. In semiconductor research, bandgap estimates are routinely used to screen large candidate spaces and guide experimental effort, making their accuracy particularly consequential. This is especially true for low bandgap materials, where small prediction errors can lead to qualitatively different assessments of device viability \cite{Dey2014Informatics, Crowley2016Resolution}. However, accurately determining bandgaps using computational methods remains a significant challenge.

Theoretical methods such as density functional theory (DFT) are commonly used but often underestimate bandgaps compared with experiments \cite{Perdew1983Physical, Borlido2020Exchange}. Greater accuracy in theoretical bandgap prediction generally requires significantly more computation \cite{Yang2023Range}, restricting its use at scale. Recently, machine learning methods have been increasingly adopted for data-driven bandgap prediction, with graph neural networks (GNNs) becoming a commonly used approach\cite{reiser2022graph}. Operating on DFT-relaxed crystal structures, GNNs represent materials as graphs of atoms and local environments, providing a framework for incorporating geometric and bonding information into learning pipelines. Unlike theoretical methods, trained machine learning models enable efficient bandgap prediction without expensive calculations at inference time.

\begin{table}[t]

\caption{\textbf{Comparison of evaluation dimensions covered by representative models and benchmark studies.} The columns indicate whether a study includes the following aspects:  Experimental data refers to training or evaluation using experimentally measured bandgaps rather than purely computational labels. Pretrain comparison indicates whether the study analyzes the effect of computational data pretraining or cross-fidelity transfer. Classical ML denotes comparisons between deep learning models and classical machine learning methods. Domain-based OOD refers to evaluation under domain-aware out-of-distribution splits defined by chemistry or structure. Existing work typically addresses these aspects in isolation, whereas RealMat-BaG integrates them within a unified benchmark framework. Filled circles denote explicit inclusion; open circles denote partial inclusion. Models used as baselines in this work are shown in bold.}
\centering
\small
\begin{threeparttable}
\renewcommand{\arraystretch}{1.3}

\begin{tabular}{lcccccc}
\hline
\textbf{Models} & \textbf{Year} & \textbf{Experimental data} &
\textbf{Pretrain comparison} & \textbf{Classical ML} & \textbf{Domain-based OOD} \\
\toprule

\textbf{CGCNN}\cite{Xie2018CGCNN} & 2018 &  &  &  &  \\
\rowcolor{rowgray}
MEGNet\cite{Chen2019Graph} & 2019 &  & \rdot & \rdot &  \\
Multi-fidelity MEGNet\cite{Chen2021LearningMEGNet} & 2021 & \rodot\textsuperscript{a} & \rdot &  &  \\
\rowcolor{rowgray}
\textbf{ALIGNN}\cite{Choudhary2021Atomistic} & 2021 &  &  &  &  \\
MODNet\cite{DeBreuck2021Materials} & 2021 &  &  & \rdot &  \\
\rowcolor{rowgray}
\textbf{LEFTNet}\cite{du2023leftnet} & 2023 &  &  &  &  \\
Masood et al.~\cite{Masood2023Enhancing} & 2023 & \rdot & \rdot &  &  \\
\rowcolor{rowgray}
CrysGNN\cite{Das2023Crysgnn} & 2023 &  & \rdot &  &  \\
\textbf{CHGNet}\cite{Deng2023CHGNet} & 2023 &  & \rdot &  &  \\
\rowcolor{rowgray}
\textbf{CartNet}\cite{Sole2025Cartesian} & 2025 &  & \rdot &  &  \\
CrysCo\cite{Madani2025Accelerating} & 2025 &  &  &  &  \\
\hline

\midrule
\textbf{Benchmarks} & & & & & \\
\midrule

\rowcolor{rowgray}
Matbench\cite{Dunn2020BenchmarkingMatbench} & 2020 & \rodot\textsuperscript{b} &  & \rdot &  \\
MatDeepLearn\cite{Fung2021BenchmarkingMatdeeplearn} & 2021 &  &  &  &  \\
\rowcolor{rowgray}
JARVIS-leaderboard\cite{Choudhary2024JARVIS} & 2024 &  &  & \rdot &  \\
Omee et al.\cite{Omee2024Structure} & 2024 &  &  &  & \rdot \\
\rowcolor{rowgray}
Li et al.\cite{Li2025Probing} & 2025 &  &  & \rdot & \rdot \\
RealMat-BaG (\textbf{Ours}) & 2026 & \rdot & \rdot & \rdot & \rdot \\
\hline
\bottomrule
\end{tabular}

\begin{tablenotes}[flushleft]
\footnotesize
\item[a] Multi-fidelity MEGNet requires a commercial ICSD license.
\item[b] Matbench: GNNs were not evaluated on experimental datasets.
\end{tablenotes}

\end{threeparttable}

\label{tab:review}
\end{table}
However, the lack of real-world evaluation of machine learning models arises from three key challenges: data fidelity, evaluation protocols, and model interpretability, which together obscure their reliability. The first is the data fidelity challenge between model training and experimentally measured bandgaps. Deep learning methods, including GNNs, generally require large-scale training data to learn complex structure-property relationships. However, experimental bandgap datasets remain limited and are typically curated from literature, where bandgaps are often reported only at the compound level (e.g., by chemical formula or material name). Consequently, most experimental datasets do not provide explicit machine-readable crystal structures, limiting their direct use in crystal-graph-based GNN pipelines\cite{Jha2019Enhancing, Dong2022Auto}. As a result, most models are trained and evaluated primarily on large DFT-relaxed computational datasets, which systematically underpredict bandgaps. This raises questions about how well models developed on computational benchmarks transfer to experimental bandgap prediction (Table~\ref{tab:review}) and how effectively computational pretraining can support real-world applications. These limitations motivate benchmarks based on experimental data to evaluate model reliability in practical settings.

Secondly, a challenge lies in model generalization under shifts in material space. Performance is typically reported using a random train–test split, which mainly probes interpolation within the training distribution and may overestimate generalization. In practical discovery settings, however, models are often applied to unseen materials chemically or compositionally distinct from the training data \cite{Meredig2018Can}. Recent benchmarks address this by defining out-of-distribution (OOD) splits that better simulate real discovery scenarios (Table~\ref{tab:review}). In particular, domain-based splits defined by chemistry or structure categories provide a practical way to simulate these shifts.

Moreover, the interpretability challenge limits trust and adoption of machine learning predictions in experimental workflows. For many materials-synthesis researchers, using machine learning models to guide costly synthesis and characterization raises concerns about transparency, reproducibility, and failure modes\cite{Zhong2022Explainable, Oviedo2022Interpretable}, especially for GNNs, whose complex architectures often operate as black boxes \cite{Li2022Interpretable, Joshi2021Review}. It remains unclear which elemental, bonding, or structural features different model families rely on, as existing evaluations focus more on predictive accuracy than mechanistic interpretability \cite{Gao2025GCPNet, Teng2025Atomic}.

Notably, machine learning methods with intrinsic interpretability exist \cite{Murdoch2019Definitions}. Classical machine learning methods are widely used for predicting tabular material properties\cite{Talapatra2023Band, Moeini2024Machine}. Compared with deep neural networks, they offer a more transparent mapping between input features and predictions, enabling direct element-level attribution while retaining competitive predictive performance. While GNNs could offer insights into geometric structures, classical models facilitate the quantification of the influence of elemental property atomic encoding at a different level. Moreover, their performance has been shown to rival, and in some cases surpass, deep learning models in fields such as neuroimaging, computational neuroscience, and cybersecurity\cite{He2020Deep, Engemann2022Reusable,  Mehavilla2026Evaluating}. Despite this, classical machine learning methods are rarely used as baselines for structure–property prediction (Table~\ref{tab:review}), and their potential for interpreting crystal structure information remains underexplored.

Despite rapid progress in model design and benchmarking, existing studies typically address these challenges in isolation. As summarized in Table~\ref{tab:review}, prior work rarely evaluates experimental datasets, pretraining effects, classical machine learning baselines, and domain-aware OOD splits within a unified framework. making it difficult to assess how representation learning, inductive bias, and training strategy jointly influence model reliability.

To address these challenges, we introduce RealMat-BaG, a benchmark for evaluating machine learning models under more realistic deployment scenarios. We compile an open-access experimental bandgap dataset with aligned crystal structures, enabling direct training and evaluation on structural representations, and facilitating evaluation of cross-fidelity transfer between Perdew-Burke-Ernzerhof(PBE)-level DFT and experimental data. We further provide a unified evaluation framework that compares GNN architectures, computational pretraining, and classical machine learning baselines under both random and domain-based OOD splits. Finally, we incorporate interpretability analyses using node- and edge-level gradient saliency for GNNs and SHapley Additive exPlanations (SHAP) for classical models to examine whether predictions align with chemically meaningful structure and elemental-property relationships.

Together, RealMat-BaG provides comparative insights on the progress of model development for bandgap prediction tasks and a framework for assessing accuracy, generalization, and interpretability in materials machine learning.

\begin{figure*}[t!]
    \centering
    \includegraphics[width=\textwidth]{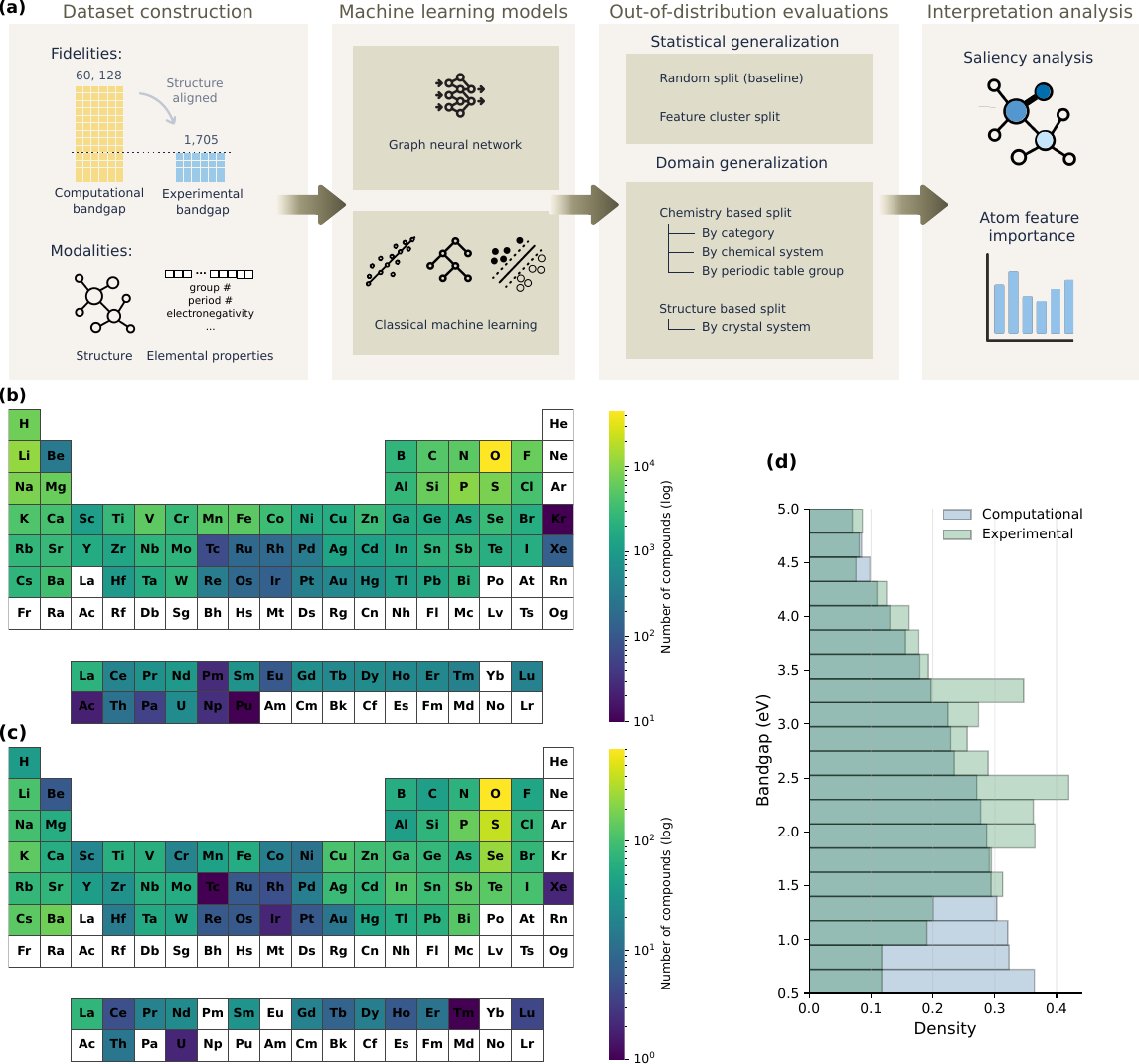}

\caption{\textbf{Overview of the RealMat-BaG benchmark design.}
\textbf{a,} Schematic overview of the benchmark pipeline. The dataset is constructed across two fidelity levels: a computational dataset derived from the Materials Project and an experimental dataset compiled from Zhuo et al. (2018)\cite{Zhuo2018Predicting}, BandgapDatabase1\cite{Dong2022Auto}, and DS2\cite{Masood2023Enhancing}. Two modalities, crystal structure and elemental property information, are incorporated for model training. Model performance is evaluated under an in-distribution random split as well as multiple out-of-distribution (OOD) settings. Interpretability is analyzed using graph-based gradient saliency methods for GNNs and SHapley Additive exPlanations (SHAP) for atomic feature importance in classical machine learning models, including linear regression, random forest regression, and support vector regression.
\textbf{b-c,} Periodic table heatmaps showing the frequency of elements in the \textbf{b,} computational and \textbf{c, }experimental datasets, where each element is counted once per compound. Color intensity indicates the number of compounds containing a given element (logarithmic scale), and lanthanides and actinides are shown separately. Elements are colored white if absent from the dataset.
\textbf{d,} Distribution of bandgap values in the computational and experimental datasets, illustrating systematic differences in bandgap value density between the two data fidelities.}

    \label{fig:fig1}
\end{figure*}

\section*{Results}

\subsection*{The RealMat-BaG benchmark framework}

The design of the RealMat-BaG benchmark aims to provide a comprehensive and practically relevant framework for evaluating machine learning models for experimental bandgap prediction. Rather than focusing on a single evaluation criterion, our benchmark integrates multiple complementary settings to reflect different aspects of model performance. By centering evaluations on experimentally measured bandgaps aligned with crystal structures, the benchmark evaluates model performance under conditions that more closely resemble real materials discovery workflows.

RealMat-BaG is organized into four complementary evaluation settings: random split, feature-based split, chemistry-based OOD splits, and structure-based OOD split. The random split and feature-based OOD split are designed from a \textit{statistical generalization} perspective. In the feature-based OOD setting, materials are partitioned according to clustering in the feature space. These settings evaluate data efficiency and robustness under statistically induced variations. The chemistry-based and structure-based OOD splits are formulated from a \textit{domain generalization} perspective and defined using physically interpretable variables. Chemistry-based splits are constructed according to material categories, chemical systems, and periodic groups, while the structure-based split is defined by crystal system. By explicitly evaluating transfer across physically meaningful material domains, these settings better reflect realistic materials discovery and deployment scenarios.

We employ gradient saliency maps and SHAP-based feature importance to examine whether model predictions rely on physically meaningful elemental and bonding motifs, providing insight into learned inductive biases. Performance is quantified using metrics such as mean relative absolute error (MRAE), which better reflects practical accuracy requirements across different bandgap regimes than mean absolute error (MAE). Together, RealMat-BaG provides a unified framework for evaluating predictive accuracy, robustness under distributional shifts, and the consistency of learned representations with atomic and bonding characteristics, offering a broader view of model reliability under practical deployment conditions.

\subsection*{Bridging the data fidelity gap with experimental data}
\begin{figure*}[t!]
    \centering
    \includegraphics[width=0.65\textwidth]{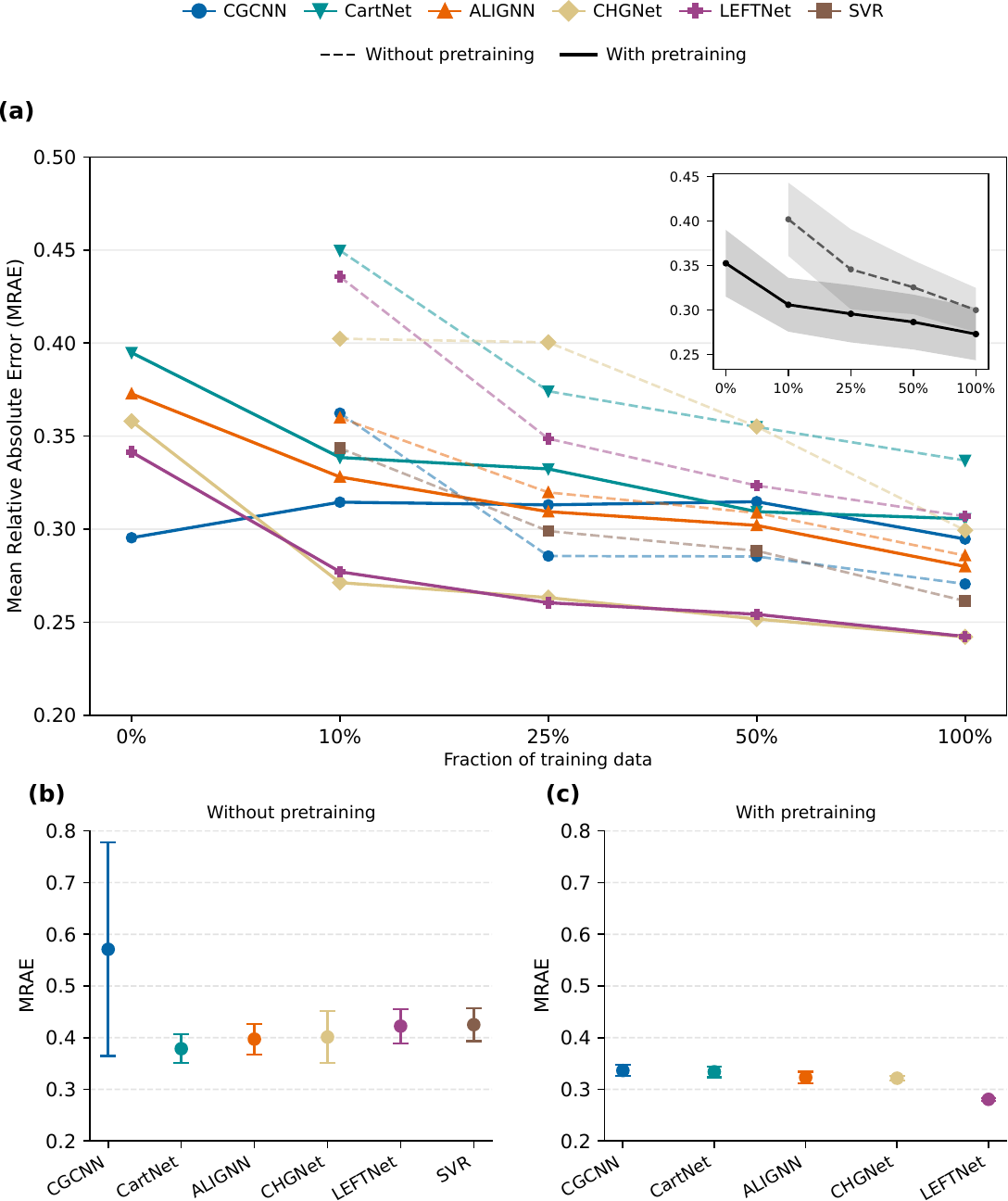}
\caption{
\textbf{Effect of computational pretraining on data efficiency and robustness in experimental bandgap prediction.}
\textbf{a, }Effect of pretraining across experimental training data fractions. Test mean relative absolute error (MRAE) of six models as a function of the fraction of available experimental training data (full training set, $n=1,534$), evaluated on a fixed hold-out test set shared across all experiments. Solid lines denote models initialized with computational pretraining (the 0\% point corresponds to direct transfer), whereas dashed lines indicate models without pretraining. The inset shows the mean performance across five GNN models, with shaded regions indicating the standard deviation over 10-fold cross-validation. SVR is not included in the inset because it does not use pretraining. Supplementary Figure 4 shows the difference in MRAE between models with and without pretraining across different training data fractions. See Supplementary Table 12 for the results of all models on the full training set.
\textbf{b-c,} Feature-based OOD performance \textbf{b,} without and \textbf{c,} with pretraining, reported as test MRAE (mean $\pm$ standard deviation) averaged over 10-fold cross-validation and evaluated on a fixed hold-out test set. This split probes robustness to distributional shifts defined in feature space.
}
\label{fig:pretrain}
\end{figure*}

Experimental datasets exhibit systematically different distributions from computational datasets, reflecting distinct scientific priorities and measurement feasibility \cite{Kumagai2022Effects}. Large-scale computational databases such as the Materials Project \cite{Jain2013CommentaryMP} sample broad regions of chemical and structural space, whereas experimental bandgap measurements are constrained by synthesis accessibility and research focus, resulting in imbalanced coverage across compound categories.

In this work, the computational dataset is constructed from the Materials Project database, which provides DFT-relaxed crystal structures and corresponding PBE bandgap values. To better align evaluation with real-world materials practice, we compiled an experimental bandgap dataset from three sources (Zhuo et al. (2018) \cite{Zhuo2018Predicting}, BandgapDatabase1 \cite{Dong2022Auto}, and DS2 \cite{Masood2023Enhancing}). As shown in Fig.~\ref{fig:fig1}b–c, the computational dataset spans a broader range of elements across the periodic table, including many transition metals and heavy elements sparsely represented in the experimental dataset. Supplementary Figure 1 further illustrates that the experimental dataset is dominated by chalcogenides, reflecting their extensive investigation due to tunable electronic properties and relevance to optoelectronic applications \cite{Woods-Robinson2020Wide}. 

Figure~\ref{fig:fig1}d shows that the two datasets also exhibit distinct bandgap distributions. Experimental measurements are concentrated in an intermediate bandgap range (2.0–2.5~eV), whereas computational data span a broader range. This mismatch in both input space and target distribution highlights a fundamental data fidelity gap between computational and experimental datasets.

\subsection*{Impact of experimental data on data efficiency and robustness}

To evaluate the impact of experimental data on mitigating the data fidelity gap, we analyze model performance on experimental bandgap measurements under varying amounts of supervision. We focus on data efficiency, transferability from computational pretraining, and robustness under feature-level distribution shifts. Pretraining on computational data is introduced to assess whether representations learned from low-fidelity supervision can adapt to experimental high-fidelity bandgap prediction. In this work, the pretraining procedure is applied to the deep learning models, while classical machine learning models are trained directly on the experimental data. We report the best classical machine learning model, support vector regression (SVR) results; results for linear regression (LR) and random forest regression (RFR) are provided in the Supplementary Section 7.

As shown in Fig.~\ref{fig:pretrain}, when only 10\% of the experimental training data is available, computationally pretrained models significantly outperform those trained from scratch. Models trained solely on computational data exhibit large errors relative to experimental bandgaps (Supplementary Table 11). As the training set size increases, the performance gap narrows, indicating reduced reliance on pretrained representations with sufficient experimental supervision.

Under the random split, models exhibit distinct trends between pretrained and non-pretrained configurations. The simpler architecture CGCNN (Fig.~\ref{fig:pretrain}a, blue line) shows only marginal benefits from pretraining; the model without pretraining outperforms its pretrained counterpart across most training fractions. Geometry-focused models such as CHGNet (Fig.~\ref{fig:pretrain}a, yellow line) and LEFTNet (Fig.~\ref{fig:pretrain}a, purple line) benefit substantially from pretraining. This suggests that geometric representations learned from lower-fidelity data facilitate experimental bandgap prediction. By comparison, CGCNN appears to gain little from pretraining, likely because the available experimental data are sufficient for this simpler architecture.

Beyond in-distribution data efficiency, we examine whether pretraining improves robustness under distribution shifts in experimental feature space. In Fig.~\ref{fig:pretrain}b-c, pretrained models generally exhibit lower MRAE and reduced variance across feature OOD splits. These results indicate improved robustness under feature-level distribution shifts, mitigating extrapolation instability. Most GNN models outperform classical machine learning methods under feature OOD, indicating stronger robustness to feature-level shifts.

\begin{figure*}[t!]
    \centering
    \includegraphics[width=0.95\textwidth]{figures/Figure_3.pdf}

\caption{\textbf{Model generalization under domain-based out-of-distribution (OOD) evaluation settings.}
\textbf{a-b,} Category leave-one-material-out (LOMO) evaluation with a shared color scale, where lower MRAE indicates better performance. Each cell reports the test MRAE on one held-out material category, and the value shown at the right of each row is the average across categories. In both panels, the material categories on the horizontal axis are ordered by sample count in descending order, and the number in parentheses indicates the number of compounds in that category. Panel \textbf{a,} shows pretrained models, whereas panel \textbf{b,} shows the corresponding models without pretraining and additionally includes SVR as a classical machine-learning baseline.
\textbf{c,} Random-split performance as an in-distribution baseline.
\textbf{d–e,} Chemistry-aware OOD generalization under \textbf{d,} chemical system and \textbf{e,} periodic group splits. \textbf{f,} Structure-aware OOD generalization under the crystal system split. Points indicate the MRAE across outer folds of 10-fold cross-validation (7-fold for the crystal system split), with error bars denoting the standard deviation.
Across all settings, performance degrades under OOD splits, with domain-dependent variations in robustness. See Supplementary Table 10 for the experimental results corresponding to panels \textbf{c–f}.}

    \label{fig:ood_splits}
\end{figure*}

\subsection*{OOD generalization across material domains}

To assess model robustness under distribution shifts arising across material domains, we evaluate performance under domain-based splits. These splits are defined by systematic changes in chemical composition or crystal structure between training and test sets, including chemistry-based splits (material categories, chemical systems, and periodic groups) and a structure-based split defined by crystal system.

Categorical leave-one-material-out (LOMO) splits reveal variation in performance across material families, indicating that category-level generalization difficulty differs substantially across categories (Fig.~\ref{fig:ood_splits}a-b). Among the evaluated categories, silicides and antimonides are generally the most challenging. This likely reflects the diverse bonding environments and structural motifs characteristic of these materials. Pretraining also leads to the largest performance degradation in these two categories. This pattern may suggest that pretrained representations inherit systematic PBE error patterns, as these two categories exhibit the largest PBE-level errors (Supplementary Table 9). It may also reflect the small sample sizes of these two categories.

Figure~\ref{fig:ood_splits}c–f compares domain-based OOD splits with the random split. The higher MRAE observed under OOD settings indicates reduced model robustness under domain shifts. CHGNet demonstrates the strongest cross-domain generalization, suggesting that richer geometric and structural inductive biases improve extrapolation. Without computational pretraining, simpler baselines (CGCNN, CartNet, and SVR) often achieve competitive performance, indicating reduced reliance on pretraining and greater stability when trained directly on limited high-fidelity data in OOD settings. 

Across the domain partitions, we observe a clear hierarchy of difficulty: chemical system splits are the easiest, whereas LOMO and periodic group splits are the most challenging. The relatively strong performance on chemical system splits likely reflects a smaller effective distribution shift, as the large number of chemical systems leads to substantial overlap in elemental composition between training and test sets.

\subsection*{Effect of atomic encoding on model generalization}

Different models adopt different atomic encodings. Motivated by chemically informed priors encoded through elemental properties (Prop) in CGCNN, we investigate whether incorporating elemental property encodings, rather than relying solely on atomic number features (Z), improves model performance and generalization. To this end, we control the atomic encoding across four GNN models and evaluate performance under random splits, feature-based OOD splits, chemistry-based LOMO splits, and structure-based crystal system splits.

\begin{figure*}[t!]
    \centering
    \includegraphics[width=\textwidth]{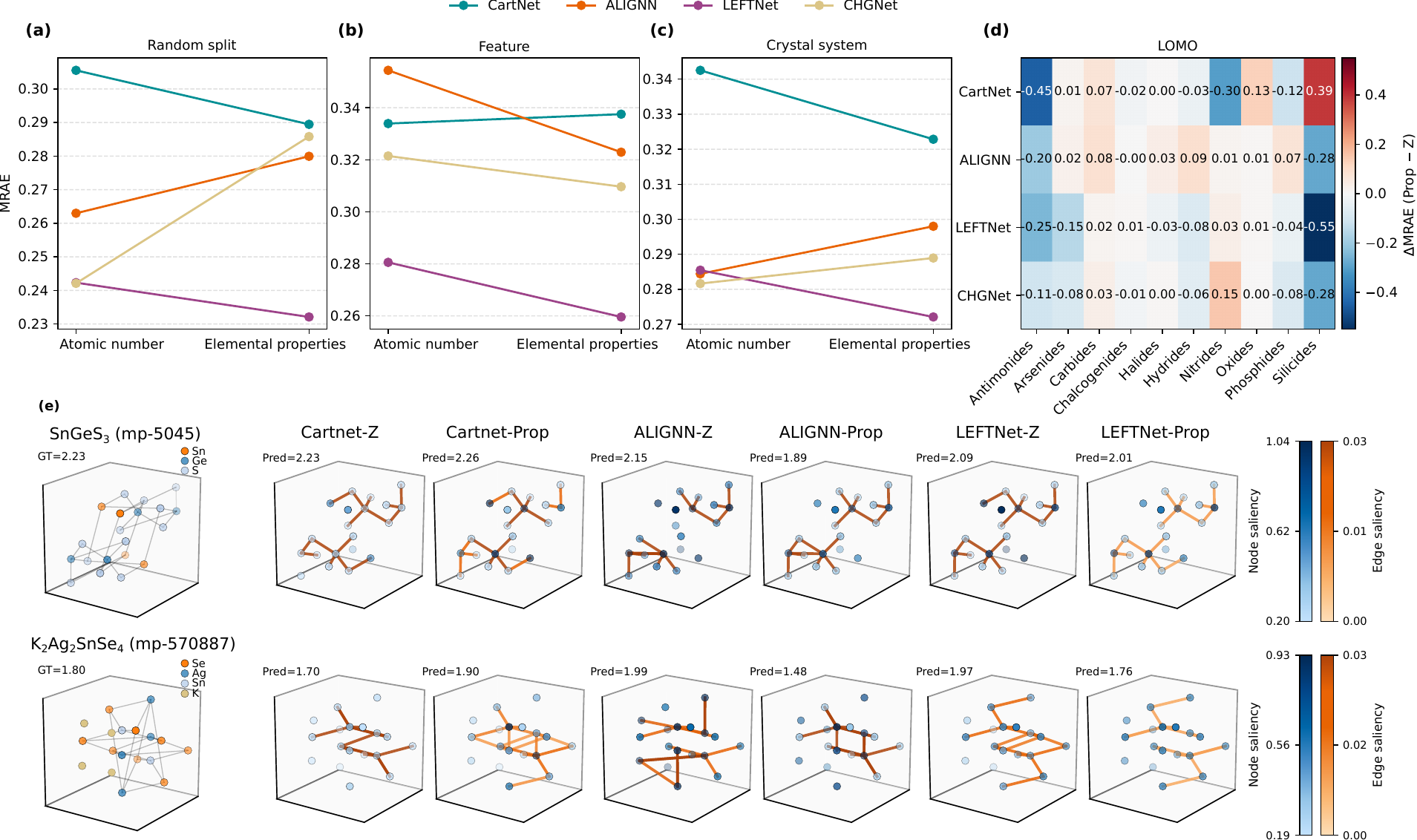}
    \caption{\textbf{Effect of atomic encoding across evaluation settings, and model interpretability at the structural level.}
     Comparison of test MRAE under atomic number (Z) and elemental property (Prop) encodings across four GNN models.
\textbf{a,} Random split.
\textbf{b,} Feature-based out-of-distribution (OOD) split.
\textbf{c,} Crystal-system-based split.
\textbf{d,} Category-level leave-one-material-out (LOMO) results, shown as a heatmap of $\Delta$MRAE (Prop $-$ Z) to provide a clearer view of patterns across models and categories. Supplementary Table 13 reports the full results for each model and category.
\textbf{e,} Node- and edge-level gradient saliency maps (atoms and interatomic bonds, respectively) for two representative materials SnGeS\textsubscript{3} (mp-5045) and K\textsubscript{2}Ag\textsubscript{2}SnSe\textsubscript{4} (mp-570887), obtained from CartNet, ALIGNN, and LEFTNet under both Z and Prop encodings. These maps provide structural interpretation by highlighting the atomic sites and bonding motifs that contribute most strongly to the predictions. Colors indicate gradient saliency scores averaged over 10 random-split folds, with node colors reflecting atomic importance and edge colors reflecting bonding importance. Ground-truth (GT) and predicted (Pred) bandgaps are shown for reference. For visualization clarity, only the top 15 most salient edges are displayed in each subfigure.}

    \label{fig:encoding_method}
\end{figure*}

Under random splits, Prop encoding improves performance for CartNet and LEFTNet, but not for ALIGNN and CHGNet (Fig.~\ref{fig:encoding_method}a). In OOD settings (Fig.~\ref{fig:encoding_method}b-c), Prop encoding yields modest improvements, with minor degradation in some cases. In the LOMO scenario, encoding benefits vary across categories and models (Fig.~\ref{fig:encoding_method}d). For the most difficult categories (antimonides and silicides), Prop encoding shows the largest improvements across models.

Overall, these results indicate that elemental property encodings are not a universal solution to distribution shift, as their benefits under feature-based and structure-based OOD settings remain limited. Prop encoding reduces the MRAE in chemistry-based LOMO splits, suggesting that chemically informed atomic representations can enhance generalization across distinct material domains rather than mitigating feature-level perturbations.

\subsection*{Interpreting structural representations in GNNs via gradient saliency}

To fill the interpretation gap between learned representations and established structure–property principles, we analyze node- and edge-level gradient saliency to connect model attributions with physically meaningful atomic and bonding environments (Fig.~\ref{fig:encoding_method}e). We select two materials well predicted across models, SnGeS\textsubscript{3} (mp-5045) and K\textsubscript{2}Ag\textsubscript{2}SnSe\textsubscript{4} (mp-570887), for qualitative comparison. CartNet, ALIGNN, and LEFTNet are chosen as representative methods with distinct inductive biases, illustrating key saliency patterns without redundant visualizations. As shown in Fig.~\ref{fig:encoding_method}e, for SnGeS\textsubscript{3} (mp-5045), chalcogen–metal bonds such as S–Sn and S–Ge receive higher saliency scores, and the corresponding sulfur and metal sites are frequently highlighted. In contrast, atoms primarily serving structural or charge-balancing roles exhibit lower saliency. This behavior aligns with established electronic-structure understanding of chalcogenide semiconductors, where band-edge states are dominated by chalcogen p orbitals hybridized with metal s/p states.

A similar pattern is observed for K\textsubscript{2}Ag\textsubscript{2}SnSe\textsubscript{4} (mp-570887) \cite{tang2021band, ye2022low}. In the better-performing models, such as CartNet and LEFTNet with property-based encoding, K atoms do not participate in any highly salient edges, consistent with their limited involvement in band-edge formation from a physical perspective. Instead, salient edges are concentrated along the Ag–Se and Sn–Se covalent backbone, reflecting the dominant role of chalcogen–metal hybridization in determining the bandgap.

Beyond material-specific trends, systematic differences emerge across model architectures: models encoding higher-order geometric information, such as three-body interactions or SE(3)-equivariant message passing, tend to assign greater saliency to bonds rather than atomic sites, reflecting a stronger emphasis on local geometric relationships.

Comparing Z and Prop encodings, Prop encoding tend to distribute saliency across a connected set of chemically relevant nodes and bonds. This suggests that incorporating elemental properties enables models to differentiate between chemically similar yet geometrically distinct bonding environments, leading to a more nuanced representation of local coordination. Although gradient saliency reflects local sensitivity of intermediate representations rather than electronic-state contributions, the cross-model consistency indicates that the learned structure–property correlations are qualitatively aligned with established physical intuition rather than arising from spurious statistical associations.

\begin{figure*}[t!]
    \centering
    \includegraphics[width=0.80\textwidth]{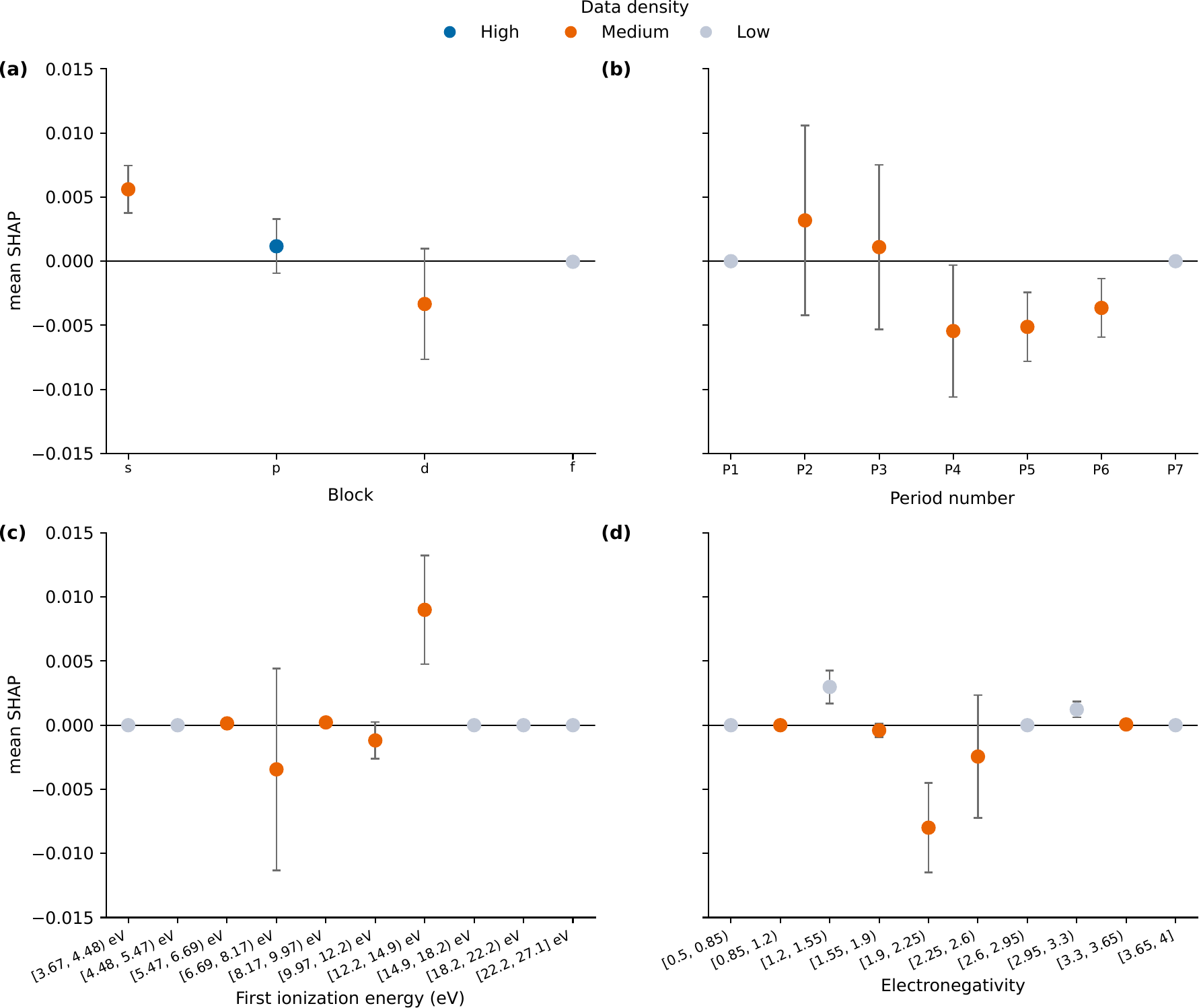}
    \caption{
\textbf{SHAP-based interpretation of elemental-property atomic encoding in SVR.} Mean signed SHAP values for four representative input properties, averaged over 10-fold cross-validation under the random split: \textbf{a,} block, \textbf{b,} period number, \textbf{c,} first ionization energy, and \textbf{d,} electronegativity. Positive SHAP values indicate that increasing the proportion of atoms within the corresponding interval increases the predicted bandgap, whereas negative values indicate the opposite trend. Error bars represent the standard deviation across folds. The color of each point indicates the data density (non-zero proportion) within the corresponding interval, categorized as low (<33\%), medium (33\%–67\%), and high (>67\%). Intervals with low density may exhibit SHAP values close to zero due to insufficient representation.}
    \label{fig:classical_ml}
\end{figure*}

\subsection*{Interpreting elemental property representations in SVR via SHAP}

While GNNs primarily provide interpretation at the structural level, classical models operate on aggregated input representations, making them more suitable for analyzing how atomic encoding influences bandgap prediction. SVR was selected for interpretation because it achieved the best overall predictive performance among the classical baselines. Figure~\ref{fig:classical_ml} presents the mean signed SHAP values of the SVR model averaged across 10 folds under the random split.

To provide a representative interpretation of model behavior, we choose four physically meaningful properties that are more relevant to bandgap from nine elemental properties used in the encoding: block, period number, first ionization energy, and electronegativity. The data density for each interval is also reported, defined as the proportion of samples with non-zero values in that interval (i.e., the non-zero rate), to account for potential sample sparsity. Intervals with low data density may yield small SHAP magnitudes due to limited representation rather than negligible feature importance. Consistently large SHAP magnitudes across folds indicate that the model robustly relies on the presence of atoms within the corresponding interval when forming bandgap predictions. A positive SHAP value suggests that increasing the proportion of atoms in that interval tends to increase the predicted bandgap, whereas a negative value indicates the opposite trend.

For the block feature (Fig.~\ref{fig:classical_ml}a), the dominant contribution of the d-block is physically consistent: a higher fraction of d-block elements is typically associated with reduced bandgaps due to the direct involvement of d orbitals in band-edge states. This suggests that the model captures a meaningful electronic-structure trend.

For period number (Fig.~\ref{fig:classical_ml}b), the observed pattern aligns with established chemical intuition: increasing the proportion of higher-period (heavier) elements generally correlates with smaller bandgaps. This behavior reflects systematic changes in valence orbital extent across the periodic table, where the more diffuse orbitals of heavier atoms tend to reduce effective orbital overlap and thus the bonding–antibonding energy splitting that contributes to the bandgap. 

Not all important patterns are straightforwardly attributable to intrinsic physical mechanisms. For example, the strong positive contribution of first ionization energy (FIE) in the 12.2–14.9 eV interval largely reflects O- and N-rich systems, where higher proportions of these elements are typically associated with larger bandgaps (Fig.~\ref{fig:classical_ml}c). The 6.69–8.17 eV range is mainly linked to d-block elements, which are more commonly found in smaller bandgap materials. Some signals may also arise from dataset composition rather than direct electronic effects. Similarly, in Fig.~\ref{fig:classical_ml}d, the electronegativity interval 2.25–2.6 likely corresponds to chalcogen-dominated systems (e.g., S/Se). If these materials systematically differ in bandgap from oxides, the associated SHAP peak may partly reflect domain imbalance rather than a purely continuous electronegativity-driven trend.

\FloatBarrier
\section*{Discussion}

RealMat-BaG evaluates the reliability of models under simulated real-world distribution shifts. While computational data cannot entirely substitute for experimental supervision, it serves as an effective resource for representation pretraining, enhancing training stability and robustness when integrated with experimental data. Pretrained models perform well when experimental data are limited or under OOD settings where the target data distribution is missing. However, pretraining can prove counterproductive if the computational data are substantially misaligned with experimental bandgaps, as systematic biases in computational datasets may propagate into the pretrained representations.

Model performance is strongly dependent on deployment scenarios. The contrasting results between random and OOD splits demonstrate that evaluation protocols relying solely on random splitting systematically overestimate a model's generalization capability in realistic applications. Domain-based splits provide a more meaningful assessment of reliability under realistic cases, when the distribution could likely shift. Consistently high MRAE, even in the relatively easy settings such as the random split (0.24–0.30), indicate that current errors are not merely numerical deviations but are significant enough to alter categorical assessments of material suitability. The poor MRAE results observed under certain LOMO categories and periodic group splits occur across different model architectures, suggesting that these generalization failures are structural rather than model-specific. For materials systems with complex bonding patterns and diverse chemical environments, existing models lack sufficient cross-element inductive capability.
Beyond predictive accuracy, interpretability analyses of structural and elemental properties reveal that both GNNs and classical models exhibit a degree of alignment with established physical intuition. This consistency suggests that the models are capturing meaningful physical representations rather than spurious statistical correlations. Notably, classical machine learning models achieve competitive performance across multiple evaluation settings and, in some cases, even outperform GNNs, while remaining computationally inexpensive and practical. SHAP analysis further reveals that these models rely on chemically consistent features. However, these models operate on averaged representations and therefore do not explicitly capture geometric relationships. This limitation is discussed further in Supplementary Section 3. Consequently, evaluating interpretability could serve as a critical validation tool to ensure model reliability in complex AI-driven material discovery tasks.

Future work should consider moving beyond the current emphasis on structure and composition by enriching the metadata associated with bandgap datasets. Bandgaps are influenced not only by crystal structure, but also by extrinsic factors such as defects, synthesis conditions, phase purity and measurement protocols. More efforts need to be devoted to collecting and standardising such metadata to make it AI-ready. Incorporating richer metadata would enable more realistic multimodal benchmarks.

\section*{Methods}

\subsection*{Dataset construction}

Our data collection and curation process has two steps:

\textbf{PBE data filtering.}
We first collected data from the Materials Project, which provides both 3D structures and PBE bandgap values. To ensure relevance to semiconductor behaviour, we excluded entries with formulas containing more than eight elements or with bandgap values outside the range of 0.5–5 eV. After this filtering process, 60,218 entries remained. This gives our computational dataset. 

\textbf{Experimental data integration.}
We assembled the experimental bandgap dataset by merging three sources:
(i) Matbench-Expt \cite{rubungo2024llmmatbench}, which compiles data from Zhuo et al. (2018) \cite{Zhuo2018Predicting};
(ii) BandgapDatabase1 \cite{Dong2022Auto}; and
(iii) DS2 \cite{Masood2023Enhancing}. 

All entries were matched to Materials Project (MP) records by reduced chemical formula. For BandgapDatabase1, we removed records lacking a DOI to ensure traceability, then matched 39{,}300 records to MP. To guarantee a single numeric target per material, we excluded entries reporting ranges (e.g., 3.0–3.2\,eV). When multiple experimental measurements existed for the same formula, we computed the median across measurements within 0.5–5\,eV to mitigate outliers and measurement inconsistencies.

When multiple MP structures (polymorphs/isomers) shared the same formula, we selected the MP entry with the lowest formation energy (0\,K) as a proxy for thermodynamic stability.
To ensure a single record per composition across sources, we applied a fixed precedence rule. If a composition appeared in BandgapDatabase1, we used that value. If it was not available there but appeared in DS2, we used the DS2 entry. If neither source provided a record, we used the value from Matbench-Expt. This ordering reflects differences in data provenance: BandgapDatabase1 provides DOI-linked experimental records, while DS2 contains curated, human-compiled values. After filtering and structure alignment, the final dataset contains 1,705 experimental samples with associated crystal structures, including 1,183 from BandgapDatabase1, 185 from DS2, and 337 from Matbench-Expt.

The curated bandgap dataset is released as an open-access resource, while the corresponding crystal structures (CIF files) can be retrieved from the Materials Project database using the matched MP identifiers.

Crystal structures were parsed from CIF files using pymatgen \cite{Ong2013Python} and represented as periodic \texttt{Structure} objects. These representations were subsequently converted into model-ready inputs using consistent neighbor definitions and geometric featurisation across all methods.

\subsection*{Data splits}
The experimental dataset is partitioned into several evaluation settings to simulate different regression tasks.

\textbf{Random split.}
The experimental dataset was first partitioned into a fine-tuning set and a test set using a 0.9:0.1 split. To ensure comparable material-category distributions between the two sets, we adopted a stratified random splitting strategy, sampling data points proportionally from each category. In total, 171 out of 1,705 materials were assigned to the test set, which was held out exclusively for final performance assessment and not used during training or validation. The remaining 1,534 materials constituted the fine-tuning set and were used for cross-validation.

Given the limited size of experimental datasets, we further investigated model performance under reduced training data regimes. Subsets of the fine-tuning set were constructed by stratified random sampling across material categories, with sizes corresponding to 10\% (153 samples), 25\% (383 samples), and 50\% (767 samples) of the full fine-tuning data. All subsets were drawn exclusively from the fine-tuning set to prevent data leakage.

\textbf{Feature-based OOD split:} out-of-distribution data points are selected using $k$-means clustering on the feature space. We identify the cluster whose centroid is farthest from the global mean and treat all materials in this cluster as OOD test samples. When using $k=8$, this procedure yields an OOD subset of 189 materials. Supplementary Figure 3 illustrates the selected cluster in a two-dimensional $t$-SNE projection.

\textbf{Leave-one-material-out (LOMO):} Each material was assigned to a single category according to the presence of specific elements in its chemical formula (see Supplementary Table 1). To avoid ambiguity and prevent label ``leakage'' across categories, we removed 451 out of the original 1,705 materials that either belonged to multiple categories or could not be assigned to any category. Supplementary Table 2 reports the number of materials per category before filtering, while the final curated dataset contains only materials that can be uniquely classified. 

\textbf{Chemical system, periodic group and crystal system splits:} To construct chemically meaningful train–test partitions, we employed MatFold \cite{Witman2025MatFold}, a cross-validation framework developed for materials science. Chemical system and periodic group splits used 10-fold cross-validation, whereas crystal-system splits used 7 folds.

All models are trained separately on each of the above splits using the training setup described in Evaluation~\ref{sec:Evaluation}.

\subsection*{Models}
\subsubsection*{Classical machine learning methods}
We consider three classical machine learning baselines: Linear Regression (LR), Support Vector Regression (SVR), and Random Forest Regression (RFR), built on fixed-length descriptors derived from elemental properties and interatomic distance encodings. Following CGCNN’s feature design, we use similar elemental and radial-basis distance features but concatenate them at the \emph{compound} rather than atomic level. Specifically, we first compute atom-level encodings $\mathbf{h}_i$ from the raw inputs and then form a single crystal descriptor by global averaging, 
\[
\bar{\mathbf{h}} \;=\; \frac{1}{N}\sum_{i=1}^{N} \mathbf{h}_i,
\]
where \(N\) is the number of atoms in the crystal. This yields a fixed-dimensional representation for each material that serves as input to LR, SVR, and RFR.
\subsubsection*{Graph neural networks}

The graph neural networks evaluated in this benchmark span the main architectural families used for crystalline materials. Graph-based methods represent each structure as an atomistic graph, where nodes correspond to atoms and edges encode interatomic geometry. Within this category, we include 2-body message-passing networks that use only distance-based interactions (CGCNN\cite{Xie2018CGCNN}, CartNet\cite{Sole2025Cartesian}) and 3-body networks that additionally incorporate angular information (ALIGNN\cite{Choudhary2021Atomistic}, CHGNet\cite{Deng2023CHGNet}). We also evaluate an E(3)-equivariant graph neural network (LEFTNet\cite{du2023leftnet}), which encodes rotational and translational symmetry through directional features. Higher-order n-body models (n>3) are not included due to their substantially higher computational cost.

\subsection*{Evaluation settings}
\label{sec:Evaluation}

\subsubsection*{Prediction accuracy}
We use mean absolute error (MAE), mean relative absolute error (MRAE), and the coefficient of determination ($R^2$).
MAE, a widely used metric in existing works, measures the average magnitude of errors between predicted and actual values, providing an intuitive interpretation of model performance.
\(R^2\) quantifies the proportion of total variation of outcomes the model explains.
MRAE normalizes the error, making it meaningful regardless of the bandgap's magnitude and allowing for fair comparisons across materials with diverse bandgap ranges. In particular, for materials with small bandgaps, a small absolute error may correspond to a large relative error, which can significantly affect their properties. Therefore, we emphasized MRAE in our experimental setting and selected the best model for each cross-validation fold based on the lowest MRAE.

\subsubsection*{Interpretation using gradient-based attribution for nodes and edges}

To interpret GNN model predictions at the atomic and bonding levels, we employ a gradient-based attribution method that computes saliency scores for nodes and edges of the crystal graph~\cite{Simonyan2014DeepInside, Pope2019Explainability}. 
To reduce sensitivity to model selection and training noise, we average saliency maps over the top 10 checkpoints selected by validation performance under the random split.

Given a trained GNN model $f_\theta$ that predicts the bandgap $\hat{y} = f_\theta(G)$ from an input graph $G = (V, E)$, we compute gradient-based attributions with respect to the initial node and edge embeddings obtained after the feature embedding layers and prior to message passing.

For a given prediction $\hat{y}$, we backpropagate gradients through the model to obtain
\[
\nabla_{\mathbf{h}_i} \hat{y} = \frac{\partial \hat{y}}{\partial \mathbf{h}_i}, \qquad 
\nabla_{\mathbf{e}_{ij}} \hat{y} = \frac{\partial \hat{y}}{\partial \mathbf{e}_{ij}},
\]
where $\mathbf{h}_i$ and $\mathbf{e}_{ij}$ denote the node and edge representations at this early representation level.
The elementwise product between each representation and its corresponding gradient,
\[
\mathbf{s}_i = \mathbf{h}_i \odot \nabla_{\mathbf{h}_i} \hat{y}, \qquad
\mathbf{s}_{ij} = \mathbf{e}_{ij} \odot \nabla_{\mathbf{e}_{ij}} \hat{y},
\]
defines the gradient $\times$ activation saliency tensors for nodes and edges, respectively.
Scalar importance scores are obtained by reducing these tensors along the feature dimension (using the $\ell_2$ norm):
\[
a_i = \|\mathbf{s}_i\|_2, \qquad a_{ij} = \|\mathbf{s}_{ij}\|_2.
\]

\subsubsection*{Interpretation of classical machine learning models via SHAP}
Taking advantage of the intrinsic interpretability of classical machine learning models, we employed SHAP (SHapley Additive exPlanations) to quantify the contribution of each feature used during the encoding stage. SHAP values were computed separately for each fold using the best-performing model selected for that fold. For SVR, we used Kernel SHAP to approximate feature attributions, while model-specific SHAP implementations were applied where applicable. To ensure robustness, SHAP values were computed across cross-validation folds and aggregated by reporting the mean and standard deviation for each feature. This aggregation captures the overall magnitude of feature importance independent of directionality. Feature-level attributions were analyzed separately for atom-level, neighbor-aggregated, and radial basis function (RBF) distance encodings, enabling systematic comparison of chemically informed descriptors across representation segments.

\section*{Data availability}

The dataset used in this benchmark is publicly available at https://github.com/Shef-AIRE/bandgap-benchmark. For each sample, the Materials Project identifier (MPID) of the corresponding crystal structure is provided. Crystal structures can be retrieved from the Materials Project database (https://materialsproject.org/).

\section*{Code availability}

The code for running the benchmark is available at https://github.com/Shef-AIRE/bandgap-benchmark. 
The leaderboard for model performance is available at https://omaib.github.io/leaderboards/realmat-bag.

\bibliography{sample}

\section*{Acknowledgements}
X.L. and H.L. disclose support for the research of this work from the EPSRC[UKRI396]. A.J.R discloses support for the research of this work from the EPSRC[EP/X039285/1]. R.D.J.O discloses support for the research of this work from the Royal Society[RG/R1/241060]. This work is also supported by donations from D. Naik and S. Naik.

\section*{Author contributions}
H.W., X.L., and H.L. conceived the project. H.W. conducted the computational experiments. A.J., A.J.R., and R.D.J.O. contributed to the interpretation and materials science validation of the results. H.W., X.L., A.J.R., and H.L. designed the figures and tables. H.W. wrote the manuscript. All the authors were involved in the review and editing process. 
\clearpage
\setcounter{figure}{0}
\renewcommand{\thefigure}{S\arabic{figure}}
\setcounter{table}{0}
\renewcommand{\thetable}{S\arabic{table}}
\setcounter{section}{0}
\renewcommand{\thesection}{S\arabic{section}}
\setcounter{subsection}{0}
\renewcommand{\thesubsection}{S\arabic{section}.\arabic{subsection}}
\def\mergedintomain{1}
\ifdefined\mergedintomain
\else
\PassOptionsToPackage{table}{xcolor}
\documentclass[fleqn,10pt]{wlscirep}
\usepackage[utf8]{inputenc}
\usepackage[T1]{fontenc}

\usepackage{graphicx}
\usepackage{subcaption} 
\usepackage[table]{xcolor} 
\usepackage{multirow}
\usepackage{lettrine}
\usepackage{enumitem}

\usepackage{float}
\usepackage{caption}
\usepackage{amsmath}
\usepackage{hyperref}
\usepackage{lettrine}

\usepackage{fancyhdr}
\usepackage{refcount}

\usepackage{algorithm}
\usepackage{algorithmic}
\usepackage{amsmath}
\usepackage{amssymb}
\usepackage{todonotes}
\usepackage{rotating}
\usepackage{lineno}
\usepackage{pdfpages}
\usepackage{booktabs}
\usepackage{array}
\usepackage[table]{xcolor}
\usepackage{tikz}
\usepackage{threeparttable}

\usepackage{placeins}

\hypersetup{
    colorlinks=true,
    linkcolor=black,
    urlcolor=black
    }

\setlist[itemize]{noitemsep}

\definecolor{mygray}{gray}{0.95}
\definecolor{myblue}{rgb}{0, 0, 0.99}

\definecolor{headgray}{gray}{0.90}
\definecolor{rowgray}{gray}{0.95}
\definecolor{dotred}{RGB}{222,85,82}

\newcommand{\rdot}{\tikz[baseline=-0.6ex]\fill[dotred] (0,0) circle (1.2mm);}
\newcommand{\rodot}{\tikz[baseline=-0.6ex]\draw[dotred,line width=0.5pt] (0,0) circle (1.2mm);}
\fi

\ifdefined\mergedintomain
\else
\begin{document}
\fi

\flushbottom

\section*{\begin{Large}Supplementary Information \end{Large}}

\vspace{0.2cm}

\tableofcontents
\thispagestyle{empty}

\setcounter{figure}{0}

\setcounter{table}{0}

\setcounter{section}{0}
\renewcommand{\thesection}{S\arabic{section}}

\setcounter{subsection}{0}
\renewcommand{\thesubsection}{S\arabic{section}.\arabic{subsection}}

\section{Experimental dataset evaluation}

\begin{figure*}[htbp]
    \centering
    \includegraphics[width=0.8\textwidth]{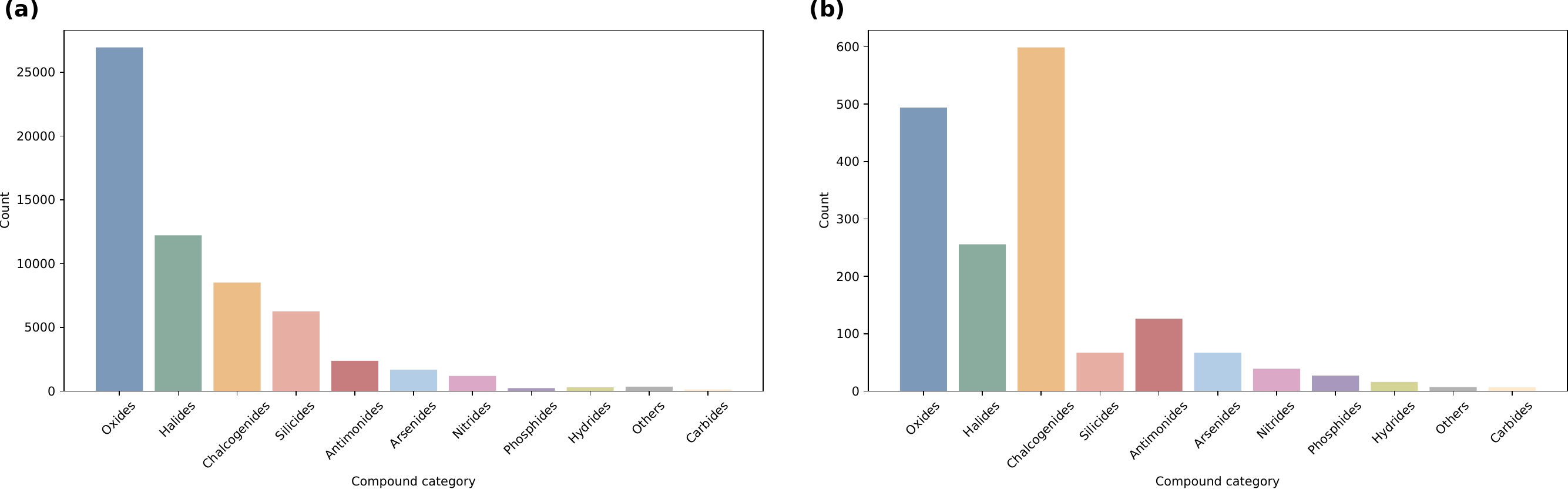}
    \caption{
    \textbf{a,} Number of materials in the computational dataset for each material category.
    \textbf{b,} Number of materials in the experimental dataset for each material category.
    }
    \label{fig:comp_exp_data_by_category}
\end{figure*}

Supplementary Figure~\ref{fig:comp_exp_data_by_category} is a supplement to Fig. 1b–d. It further shows the distributions of computational PBE and experimental bandgap data across material categories used in this paper. The comparison indicates that the experimental dataset is dominated by chalcogenides, which have been actively investigated as wide-bandgap semiconductors due to their tunable electronic properties and relevance to optoelectronic applications. As a result, models trained primarily on computational data may learn biased representations when applied without sufficient experimental supervision. The data are categorized according to the standard in Supplementary Table~\ref{tab:classification}.

\renewcommand{\figurename}{Supplementary Figure}

\renewcommand{\tablename}{Supplementary Table}
\begin{table}[htbp]
    \centering
    \caption{Categorization of materials based on chemical composition.}
\scalebox{1}{
\setlength{\tabcolsep}{1.2mm}{
\begin{tabular}{l|ccccccccccc}
        \toprule
        \textbf{Category} & Arsenides & Antimonides & Silicides & Halides & Chalcogenides & Oxides \\
        \midrule
        \textbf{Key element(s)} & As & Sb & Si & F, Cl, Br, I & S, Se, Te & O \\
        \midrule
        \midrule
        \textbf{Category} & Nitrides & Phosphides & Carbides & Hydrides & Others\\
        \midrule
        \textbf{Key element(s)} & N & P & C & H & Not classified above \\
        \bottomrule

        \hline
    \end{tabular}}}
    \label{tab:classification}
\vspace{-10pt}
\end{table}

\section{Category based Leave-one-material-out split}
The LOMO split uses the categorization scheme defined in Supplementary Table~\ref{tab:classification}. Compounds that belong to multiple categories are removed to ensure that each split is mutually exclusive and to prevent distribution leakage. As shown in Supplementary Table~\ref{tab:category_count}, only 1254 crystals are retained for the LOMO setting. Supplementary Table~\ref{tab:compound_counts} further summarizes the number of compounds in each category.

\begin{table}[hbtp]
\centering
\caption{Category distribution showing the number of categories each compound belongs to, including those that do not belong to any category.}
\label{tab:category_count}
\begin{tabular}{l|cccc}
\toprule
\textbf{Category \#} & 1 & 2 & 3 & Not belong to any \\
\midrule
\textbf{Count} & 1254 & 406 & 38 & 7 \\
\bottomrule
\end{tabular}
\end{table}

\begin{table}[hbtp]
\centering
\caption{Number of compounds in each category.}
\label{tab:compound_counts}
\begin{tabular}{l|ccccc}
\toprule
\textbf{Category} & Chalcogenides & Oxides & Halides & Nitrides & Phosphides \\
\midrule
\textbf{Count} & 486 & 481 & 159 & 36 & 27 \\
\midrule
\midrule
\textbf{Category} & Antimonides & Hydrides & Arsenides & Silicides & Carbides \\
\midrule
\textbf{Count} & 18 & 16 & 16 & 8 & 7 \\
\bottomrule
\end{tabular}
\vspace{-10pt}
\end{table}

\section{Analysis of feature space}

\begin{figure*}[htbp!]
    \centering
    \includegraphics[width=0.45\textwidth]{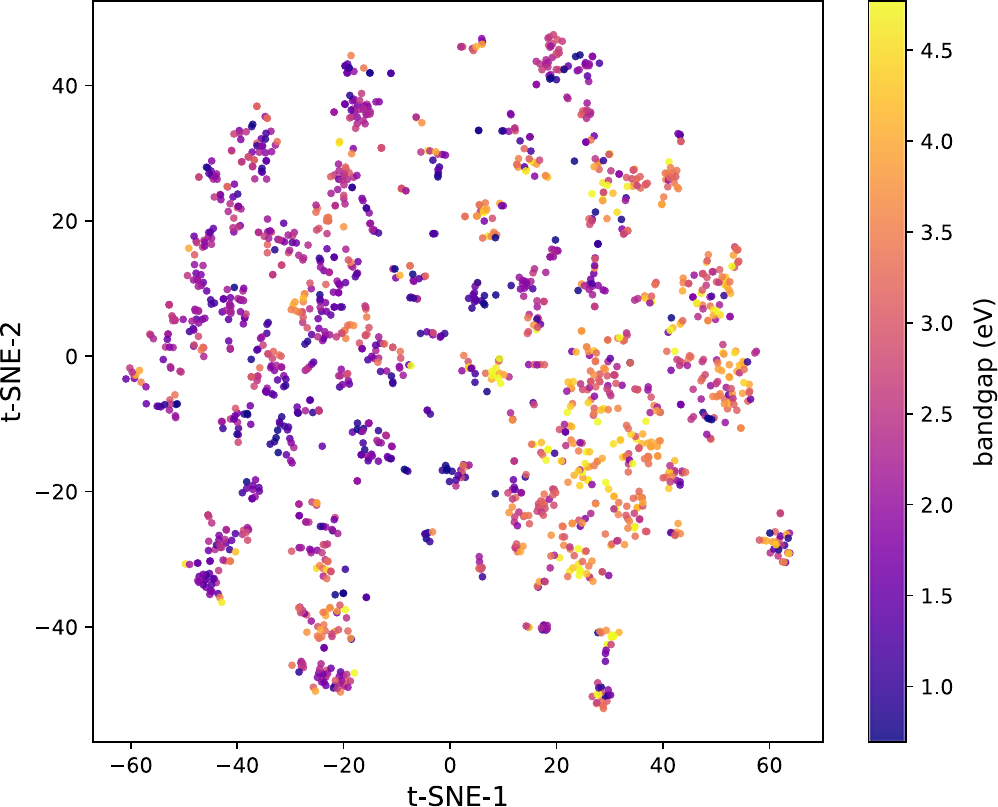}
    \caption{
        $t$-SNE visualization of the CGCNN-derived feature representation, with points colored by bandgap values.
    }

    \label{fig:tsne_band_gap}
\end{figure*}

In this paper, we adopt the fixed-dimensional material representations extracted from CGCNN as the feature space. To examine how the target property is organized under this representation, we analyze the distribution of experimental bandgap values in the corresponding $t$-SNE embedding. From Supplementary Figure~\ref{fig:tsne_band_gap}, we observe that the bandgap values vary smoothly across the embedded space, with higher bandgaps predominantly located on one side of the manifold and lower bandgaps on the opposite side.

Although the representation method employed does not explicitly encode detailed geometric or topological information, it captures the statistical frequency of different atom species and coarse-grained interatomic distance information. These factors are well known to play a dominant role in determining the overall electronic properties of materials and can significantly limit the range of bandgap values.

\begin{figure*}[htbp]
    \centering
    \includegraphics[width=0.45\textwidth]{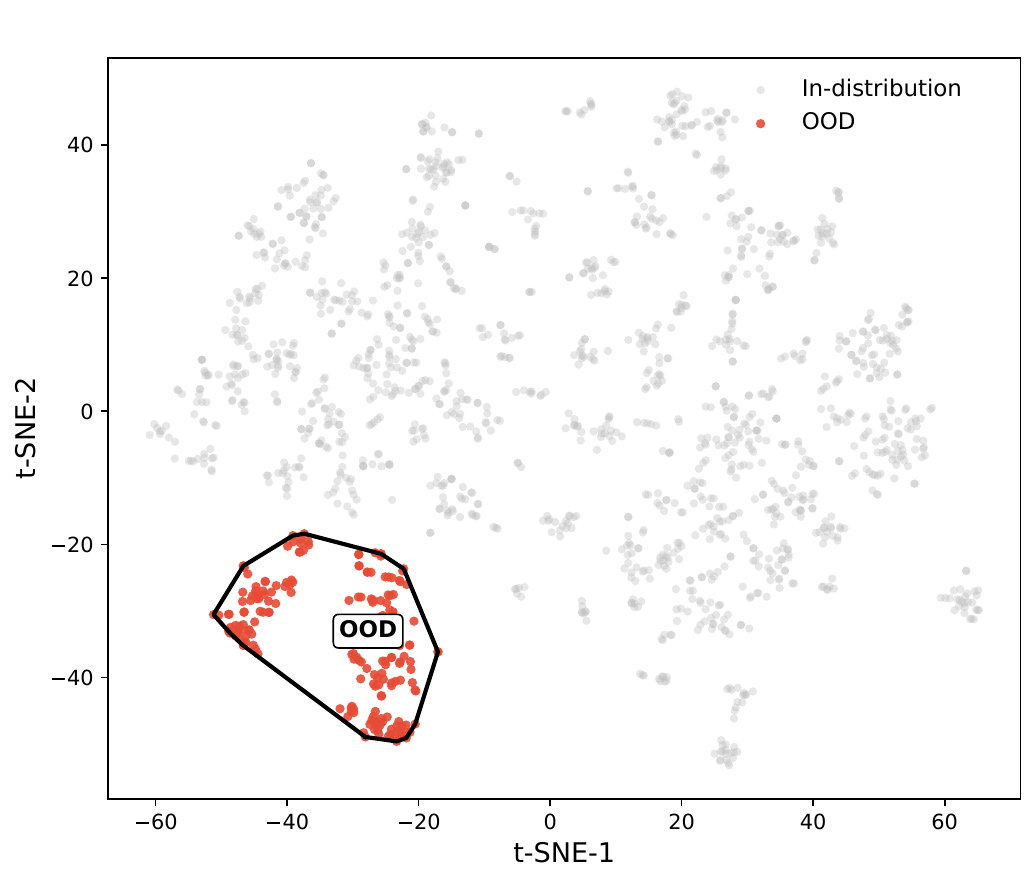}
    \caption{
        $t$-SNE projection of crystal graph embeddings from the experimental dataset. An isolated cluster of materials, automatically selected using density-based clustering, is highlighted in red and designated as out-of-distribution (OOD), while the remaining samples form the in-distribution region (grey).
    }

    \label{fig:ood_cluster}
\end{figure*}

\section{Experimental setup}

All deep models were trained using stochastic gradient descent (SGD) with a learning rate of $10^{-3}$, momentum of 0.9, and a batch size of 64. A fixed random seed (42) was used for reproducibility. Model checkpointing was performed based on the lowest validation MRE. Pretraining runs used 200 epochs, while all downstream tasks were trained for 100 epochs. For each data split or fold, models were re-initialized and trained independently on a single GPU when available. The architecture and configurations of the GNN models are summarised in Supplementary Table~\ref{tab:gnn-config}. Hyperparameter tuning is applied to the classical machine learning models, with details provided in Supplementary Table~\ref{tab:classical-hparam-random}.

\begin{table*}[t]
\centering
\caption{Key architectural and graph construction configurations of the GNN models used in this study.}
\label{tab:gnn-config}
\begin{tabular}{llr}
\toprule
Model & Configuration & Value \\
\midrule
\multirow{4}{*}{CGCNN}
& Atom embedding dimension & 64 \\
& Hidden dimension & 128 \\
& Number of convolution layers & 3 \\
& Number of output layers & 1 \\
\midrule
\multirow{6}{*}{ALIGNN}
& Hidden dimension & 128 \\
& Number of ALIGNN layers & 4 \\
& Number of GCN layers & 4 \\
& Number of Gaussian basis functions & 40 \\
& Atom cutoff radius (\AA) & 6.0 \\
& Bond cutoff radius (\AA) & 3.0 \\
\midrule
\multirow{6}{*}{CHGNet}
& Atom feature dimension & 64 \\
& Bond feature dimension & 64 \\
& Angle feature dimension & 64 \\
& Number of convolution layers & 4 \\
& Radial basis size & 31 \\
& Angular basis size & 31 \\
\midrule
\multirow{2}{*}{LEFTNet}
& Hidden channels & 128 \\
& Number of layers & 4 \\
\midrule
\multirow{4}{*}{CartNet}
& Input dimension & 256 \\
& Radial basis dimension & 64 \\
& Number of layers & 4 \\
& Maximum neighbors & 12 \\
& Cutoff radius (\AA) & 8.0 \\
\bottomrule
\end{tabular}
\end{table*}
\begin{table}[t]
\centering
\caption{Hyperparameter search space for classical machine learning models on the random split task. 
Bold values indicate the configuration selected based on the lowest validation MRAE and used for reporting test performance. Linear regression has no tunable hyperparameters under our setup and is therefore omitted from this table.}
\label{tab:classical-hparam-random}
\begin{tabular}{lll}
\toprule
Model & Hyperparameter & Candidate values \\
\midrule
\multirow{4}{*}{SVR} 
& Regularization $C$ & 1, \textbf{10}, 100 \\
& RBF kernel width $\gamma$ & 0.01, \textbf{0.1}, 1 \\
& $\epsilon$-insensitive loss & 0.01, \textbf{0.1}, 0.5 \\
& Kernel type & \textbf{rbf}, linear, poly, sigmoid \\
\midrule
\multirow{4}{*}{Random Forest} 
& Number of trees & 200, 400, \textbf{800} \\
& Maximum depth & \textbf{None}, 10, 20 \\
& Min. samples for split & \textbf{2}, 4, 8 \\
& Min. samples per leaf & \textbf{1}, 2, 4 \\
\bottomrule
\end{tabular}
\end{table}

\section{Experimental results on random splits}

\begin{figure*}[htbp!]
    \centering
    \includegraphics[width=0.8\textwidth]{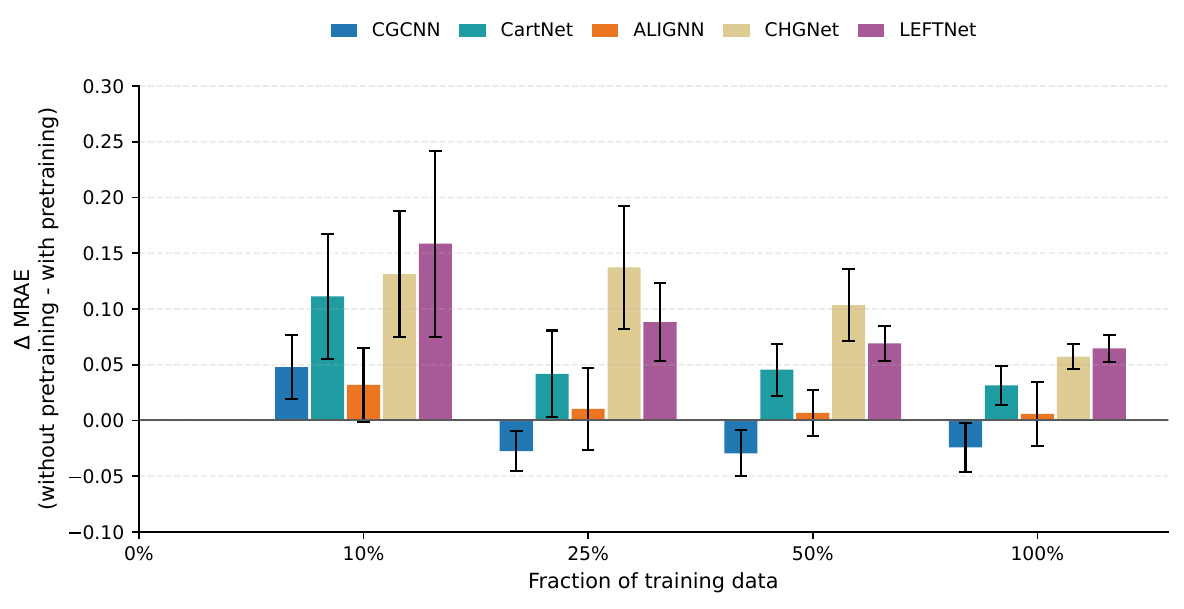}
    \caption{\textbf{Effect of pretraining across different data fractions.} 
    Difference in MRAE between models trained without and with computational pretraining across different experimental data fractions. Positive values indicate improved performance from pretraining. Error bars denote propagated standard deviation.}
    \label{fig:pretrain_supplementary}
\end{figure*}

To more directly quantify the impact of computational pretraining, Supplementary Figure~\ref{fig:pretrain_supplementary} presents the difference in MRAE between models trained without and with pretraining across different fractions of the experimental training data. The metric is defined as $\Delta$MRAE = MRAE (without pretraining) $-$ MRAE (with pretraining), such that positive values indicate performance improvements from pretraining, whereas negative values indicate performance degradation. This representation provides a complementary view to Fig. 2, highlighting how the benefit of pretraining varies across architectures and training data regimes.

\section{Experimental results on domain-based splits}

Across the LOMO split (Supplementary Table~\ref{tab:lomo_heatmap_mrae}, Supplementary Table~\ref{tab:lomo_heatmap_mae} and Supplementary Table~\ref{tab:lomo_heatmap_r2}), silicides and antimonides show larger prediction errors than others for nearly all architectures and metrics. These indicate poor generalization when these material families are excluded from training. This trend is further illustrated by the error heatmaps in Supplementary Figure~\ref{fig:heatmap_abs_error_antimonides_pretraining} and Supplementary Figure~\ref{fig:heatmap_abs_error_silicides_pretraining}, which visualize prediction errors for the two most challenging categories.

The results also reveal architecture-dependent robustness under category-level distribution shift. For instance, ALIGNN maintains relatively stable MRAE across most categories in the pretraining setting (generally $\sim$0.3–0.5), whereas CGCNN and LEFTNet show larger performance variance across chemical families.

Pretraining does not consistently improve performance under LOMO split. In the difficult categories (silicides and antimonides), pretraining sometimes increases error. For example, CGCNN MRAE on silicides increases from 0.867 (without pretraining) to 2.661 (with pretraining), and LEFTNet shows similarly large silicide errors in both settings.

\begin{figure*}[htbp!]
    \centering
    \includegraphics[width=0.8\textwidth]{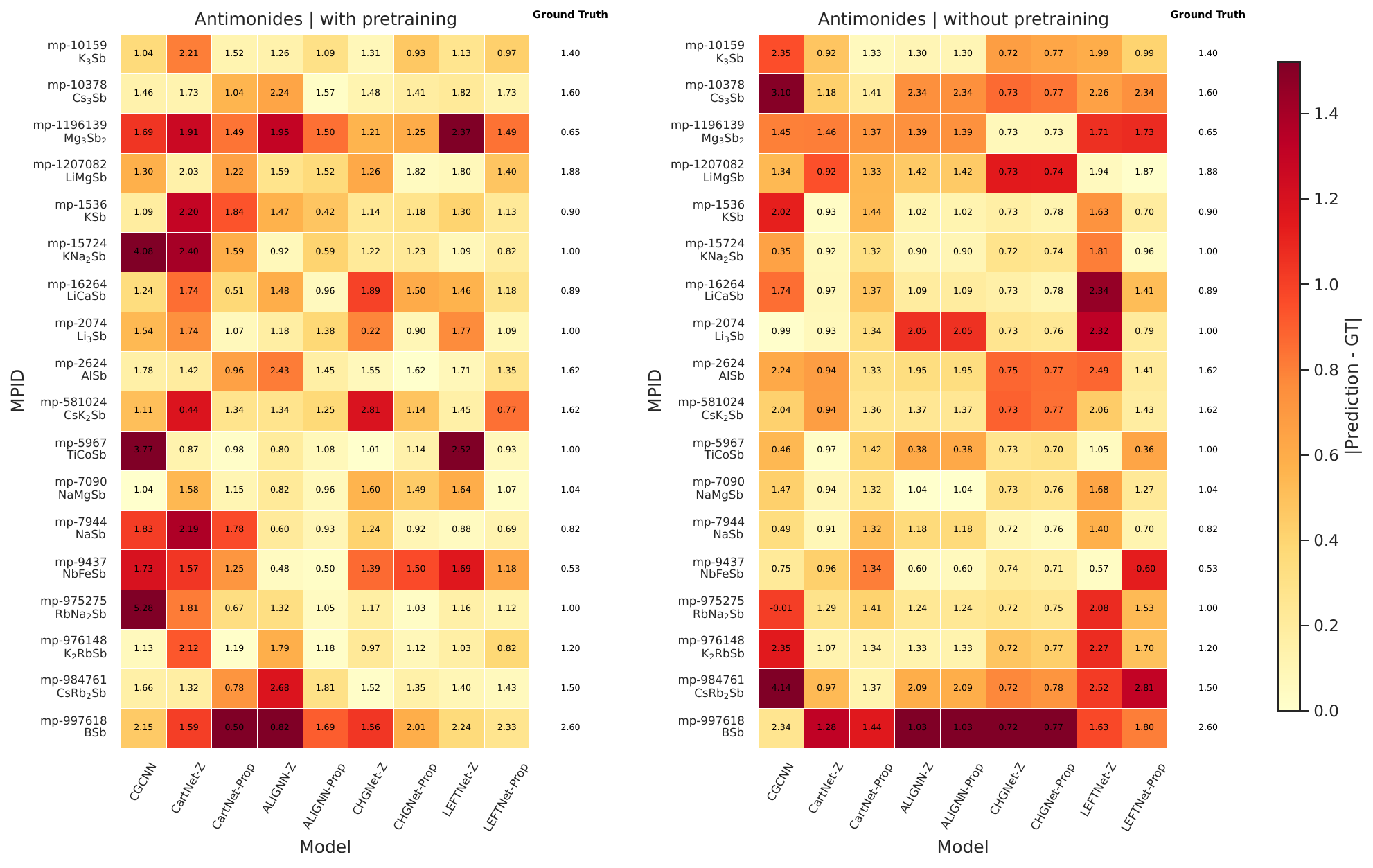}
     \caption{Heatmap of bandgap prediction errors for the antimonides split. 
    The left and right panels correspond to models with and without computational pretraining, respectively. Color indicates the absolute error $| \mathrm{Prediction} - \mathrm{GT} |$, where GT denotes the ground-truth bandgap. 
    Each cell is annotated with the predicted bandgap, and the rightmost column lists the ground-truth bandgap for each Materials Project ID (MPID).}
    \label{fig:heatmap_abs_error_antimonides_pretraining}
\end{figure*}

\begin{figure*}[htbp!]
    \centering
    \includegraphics[width=0.8\textwidth]{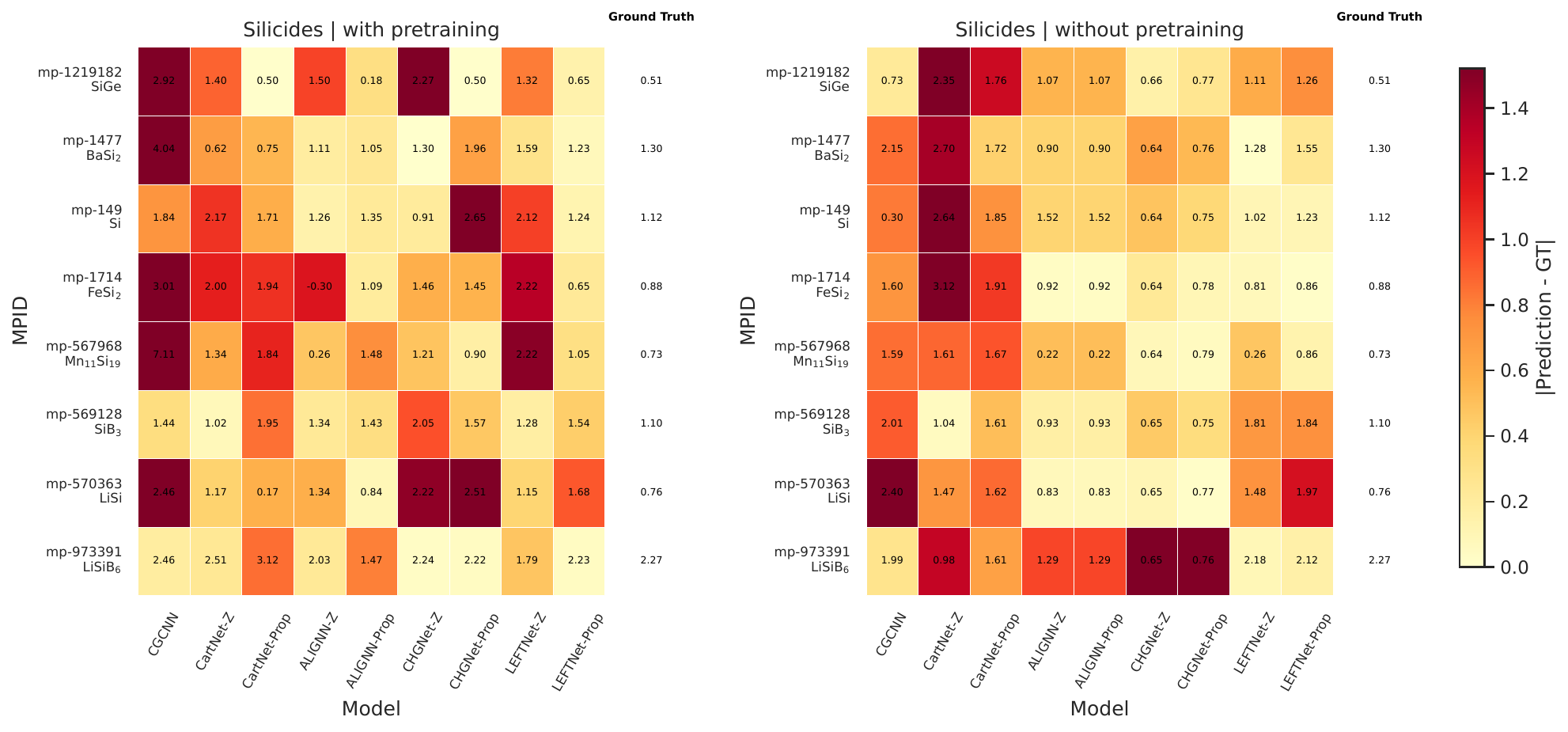}
     \caption{Heatmap of bandgap prediction errors for the silicides split. 
    The left and right panels correspond to models with and without computational pretraining, respectively. Color indicates the absolute error $| \mathrm{Prediction} - \mathrm{GT} |$, where GT denotes the ground-truth bandgap. 
    Each cell is annotated with the predicted bandgap, and the rightmost column lists the ground-truth bandgap for each Materials Project ID (MPID).}
    \label{fig:heatmap_abs_error_silicides_pretraining}
\end{figure*}

\begin{sidewaystable}
\centering
\caption{MRAE scores of each model on the LOMO split. For each category, that category is left out as the test set. Standard deviation is not reported because this split uses a single fold. All values are reported to three decimal places.}

\label{tab:lomo_heatmap_mrae}

\setlength{\tabcolsep}{4pt}
\renewcommand{\arraystretch}{1.15}

{\bfseries With pretraining}

\vspace{3mm}

\begin{tabular}{lcccccccccc}
\toprule
Model & Antimonides & Arsenides & Carbides & Chalcogenides & Halides & Hydrides & Nitrides & Oxides & Phosphides & Silicides \\
\midrule
CGCNN & 0.988 & 0.502 & 0.590 & 0.329 & 0.391 & 0.357 & 0.447 & 0.493 & 0.375 & 2.661 \\
CartNet & 0.716 & 0.401 & 0.250 & 0.318 & 0.422 & 0.542 & 0.824 & 0.371 & 0.435 & 0.619 \\
ALIGNN & 0.448 & 0.358 & 0.190 & 0.312 & 0.367 & 0.307 & 0.496 & 0.332 & 0.324 & 0.664 \\
CHGNet & 0.443 & 0.470 & 0.236 & 0.279 & 0.410 & 0.355 & 0.444 & 0.287 & 0.369 & 0.972 \\
LEFTNet & 0.555 & 0.498 & 0.255 & 0.269 & 0.363 & 0.299 & 0.488 & 0.286 & 0.369 & 0.896 \\
\bottomrule
\end{tabular}

\vspace{8mm}

{\bfseries Without pretraining}

\vspace{3mm}

\begin{tabular}{lcccccccccc}
\toprule
Model & Antimonides & Arsenides & Carbides & Chalcogenides & Halides & Hydrides & Nitrides & Oxides & Phosphides & Silicides \\
\midrule
CGCNN & 0.680 & 0.546 & 0.214 & 0.397 & 0.405 & 0.375 & 0.535 & 0.342 & 0.430 & 0.867 \\
CartNet & 0.387 & 0.358 & 0.106 & 0.365 & 0.370 & 0.379 & 0.467 & 0.351 & 0.323 & 1.072 \\
ALIGNN & 0.350 & 0.289 & 0.207 & 0.342 & 0.449 & 0.392 & 0.408 & 0.314 & 0.376 & 0.400 \\
CHGNet & 0.376 & 0.426 & 0.157 & 0.372 & 0.385 & 0.427 & 0.468 & 0.344 & 0.388 & 0.360 \\
LEFTNet & 0.687 & 0.620 & 0.199 & 0.447 & 0.441 & 0.395 & 0.465 & 0.334 & 0.291 & 0.457 \\
LR & 0.805 & 0.708 & 0.373 & 0.409 & 0.485 & 1.013 & 0.561 & 0.354 & 0.268 & 4.084 \\
RFR & 0.792 & 0.484 & 0.243 & 0.366 & 0.440 & 0.453 & 0.717 & 0.350 & 0.371 & 1.523 \\
SVR & 0.632 & 0.629 & 0.252 & 0.399 & 0.468 & 0.427 & 0.505 & 0.321 & 0.295 & 1.759 \\
\bottomrule
\end{tabular}

\end{sidewaystable}

\begin{sidewaystable}
\centering
\caption{MAE scores of each model on the LOMO split. For each category, that category is left out as the test set. Standard deviation is not reported because this split uses a single fold. All values are reported to three decimal places.}
\label{tab:lomo_heatmap_mae}

\setlength{\tabcolsep}{4pt}
\renewcommand{\arraystretch}{1.15}

{\bfseries With pretraining}

\vspace{3mm}

\begin{tabular}{lcccccccccc}
\toprule
Model & Antimonides & Arsenides & Carbides & Chalcogenides & Halides & Hydrides & Nitrides & Oxides & Phosphides & Silicides \\
\midrule
CGCNN & $\text{0.939}$ & $\text{0.601}$ & $\text{1.463}$ & $\text{0.621}$ & $\text{0.840}$ & $\text{0.790}$ & $\text{0.969}$ & $\text{1.522}$ & $\text{0.605}$ & $\text{2.077}$ \\
CartNet & $\text{0.768}$ & $\text{0.517}$ & $\text{0.635}$ & $\text{0.601}$ & $\text{0.879}$ & $\text{0.966}$ & $\text{1.295}$ & $\text{1.040}$ & $\text{0.637}$ & $\text{0.573}$ \\
ALIGNN & $\text{0.528}$ & $\text{0.449}$ & $\text{0.499}$ & $\text{0.593}$ & $\text{0.899}$ & $\text{0.827}$ & $\text{0.870}$ & $\text{0.862}$ & $\text{0.526}$ & $\text{0.505}$ \\
LEFTNet & $\text{0.472}$ & $\text{0.567}$ & $\text{0.631}$ & $\text{0.511}$ & $\text{0.710}$ & $\text{0.609}$ & $\text{0.939}$ & $\text{0.728}$ & $\text{0.515}$ & $\text{0.748}$ \\
CHGNet & $\text{0.456}$ & $\text{0.505}$ & $\text{0.624}$ & $\text{0.503}$ & $\text{0.756}$ & $\text{0.606}$ & $\text{0.753}$ & $\text{0.703}$ & $\text{0.560}$ & $\text{0.685}$ \\
\bottomrule
\end{tabular}

\vspace{8mm}

{\bfseries Without pretraining}

\vspace{3mm}

\begin{tabular}{lcccccccccc}
\toprule
Model & Antimonides & Arsenides & Carbides & Chalcogenides & Halides & Hydrides & Nitrides & Oxides & Phosphides & Silicides \\
\midrule
CGCNN & $\text{0.780}$ & $\text{0.504}$ & $\text{0.555}$ & $\text{0.667}$ & $\text{1.013}$ & $\text{0.925}$ & $\text{1.003}$ & $\text{0.851}$ & $\text{0.552}$ & $\text{0.788}$ \\
CartNet & $\text{0.477}$ & $\text{0.430}$ & $\text{0.329}$ & $\text{0.726}$ & $\text{0.968}$ & $\text{1.071}$ & $\text{1.092}$ & $\text{1.003}$ & $\text{0.461}$ & $\text{0.941}$ \\
ALIGNN & $\text{0.426}$ & $\text{0.345}$ & $\text{0.526}$ & $\text{0.656}$ & $\text{1.153}$ & $\text{1.046}$ & $\text{0.827}$ & $\text{0.754}$ & $\text{0.601}$ & $\text{0.393}$ \\
LEFTNet & $\text{0.748}$ & $\text{0.616}$ & $\text{0.537}$ & $\text{0.988}$ & $\text{1.205}$ & $\text{1.073}$ & $\text{1.003}$ & $\text{0.850}$ & $\text{0.435}$ & $\text{0.349}$ \\
CHGNet & $\text{0.540}$ & $\text{0.415}$ & $\text{0.480}$ & $\text{0.644}$ & $\text{0.932}$ & $\text{1.239}$ & $\text{1.181}$ & $\text{0.945}$ & $\text{0.598}$ & $\text{0.474}$ \\
\bottomrule
\end{tabular}

\end{sidewaystable}

\begin{sidewaystable}
\centering
\caption{$R^2$ scores of each model on the LOMO split. For each category, that category is left out as the test set. Standard deviation is not reported because this split uses a single fold. All values are reported to three decimal places.}

\setlength{\tabcolsep}{4pt}
\renewcommand{\arraystretch}{1.15}

{\bfseries With pretraining}

\vspace{3mm}

\begin{tabular}{lcccccccccc}
\toprule
Model & Antimonides & Arsenides & Carbides & Chalcogenides & Halides & Hydrides & Nitrides & Oxides & Phosphides & Silicides \\
\midrule
CGCNN & $\text{-8.422}$ & $\text{-1.559}$ & $\text{-6.931}$ & $\text{-0.063}$ & $\text{-0.263}$ & $\text{0.284}$ & $\text{-0.122}$ & $\text{-2.758}$ & $\text{-0.122}$ & $\text{-29.173}$ \\
CartNet & $\text{-2.259}$ & $\text{-1.617}$ & $\text{-0.259}$ & $\text{0.113}$ & $\text{-0.441}$ & $\text{-0.157}$ & $\text{-2.432}$ & $\text{-0.655}$ & $\text{-0.129}$ & $\text{-0.970}$ \\
ALIGNN & $\text{-1.045}$ & $\text{-0.637}$ & $\text{0.026}$ & $\text{0.041}$ & $\text{-0.487}$ & $\text{0.106}$ & $\text{0.152}$ & $\text{-0.205}$ & $\text{0.098}$ & $\text{-0.514}$ \\
CHGNet & $\text{-0.452}$ & $\text{-1.564}$ & $\text{-0.269}$ & $\text{0.319}$ & $\text{-0.064}$ & $\text{0.524}$ & $\text{0.230}$ & $\text{0.065}$ & $\text{-0.034}$ & $\text{-2.279}$ \\
LEFTNet & $\text{-0.959}$ & $\text{-1.207}$ & $\text{-0.223}$ & $\text{0.282}$ & $\text{0.106}$ & $\text{0.619}$ & $\text{-0.149}$ & $\text{0.072}$ & $\text{0.018}$ & $\text{-1.997}$ \\
\bottomrule
\end{tabular}

\vspace{8mm}

{\bfseries Without pretraining}

\vspace{3mm}

\begin{tabular}{lcccccccccc}
\toprule
Model & Antimonides & Arsenides & Carbides & Chalcogenides & Halides & Hydrides & Nitrides & Oxides & Phosphides & Silicides \\
\midrule
CGCNN & $\text{-2.969}$ & $\text{-1.651}$ & $\text{0.000}$ & $\text{-0.131}$ & $\text{-0.756}$ & $\text{0.002}$ & $\text{-0.086}$ & $\text{-0.163}$ & $\text{0.028}$ & $\text{-2.059}$ \\
CartNet & $\text{-0.621}$ & $\text{-0.363}$ & $\text{0.379}$ & $\text{-0.353}$ & $\text{-0.749}$ & $\text{-0.433}$ & $\text{-0.567}$ & $\text{-0.568}$ & $\text{0.327}$ & $\text{-3.952}$ \\
ALIGNN & $\text{-0.414}$ & $\text{0.237}$ & $\text{-0.061}$ & $\text{-0.169}$ & $\text{-1.223}$ & $\text{-0.224}$ & $\text{0.081}$ & $\text{0.044}$ & $\text{0.011}$ & $\text{0.078}$ \\
CHGNet & $\text{-1.083}$ & $\text{-0.260}$ & $\text{-0.278}$ & $\text{-0.072}$ & $\text{-0.516}$ & $\text{-0.720}$ & $\text{-0.655}$ & $\text{-0.424}$ & $\text{-0.081}$ & $\text{-0.742}$ \\
LEFTNet & $\text{-2.017}$ & $\text{-1.821}$ & $\text{-0.436}$ & $\text{-1.314}$ & $\text{-1.547}$ & $\text{-0.308}$ & $\text{-0.138}$ & $\text{-0.172}$ & $\text{0.327}$ & $\text{0.206}$ \\
LR & $\text{-2.867}$ & $\text{-2.823}$ & $\text{-1.118}$ & $\text{0.062}$ & $\text{-1.106}$ & $\text{-8.971}$ & $\text{-0.759}$ & $\text{0.065}$ & $\text{0.274}$ & $\text{-55.401}$ \\
RFR & $\text{-1.998}$ & $\text{-0.927}$ & $\text{-0.080}$ & $\text{0.050}$ & $\text{-0.383}$ & $\text{0.052}$ & $\text{-0.162}$ & $\text{-0.622}$ & $\text{0.240}$ & $\text{-5.759}$ \\
SVR & $\text{-2.079}$ & $\text{-1.914}$ & $\text{-0.304}$ & $\text{-0.551}$ & $\text{-1.341}$ & $\text{0.044}$ & $\text{0.078}$ & $\text{-0.054}$ & $\text{0.275}$ & $\text{-8.261}$ \\
\bottomrule
\end{tabular}

\label{tab:lomo_heatmap_r2}
\end{sidewaystable}

\begin{table}[t]
\centering
\caption{Average MRAE under the LOMO split by category, comparing PBE computational errors with GNN models (excluding classical machine learning models) trained with pretraining and without pretraining.}

\begin{tabular}{lccc}
\toprule
Category & PBE & Models with pretraining & Models without pretraining \\
\midrule
Antimonides & 1.358 & 0.590 & 0.496 \\
Arsenides & 1.157 & 0.450 & 0.448 \\
Carbides & 0.250 & 0.321 & 0.177 \\
Chalcogenides & 0.502 & 0.301 & 0.384 \\
Halides & 0.440 & 0.398 & 0.410 \\
Hydrides & 0.484 & 0.389 & 0.394 \\
Nitrides & 0.633 & 0.542 & 0.469 \\
Oxides & 0.340 & 0.355 & 0.337 \\
Phosphides & 0.793 & 0.388 & 0.362 \\
Silicides & 1.770 & 1.106 & 0.631 \\
\bottomrule
\end{tabular}
\label{tab:lomo_pbe_vs_pretrain_vs_no_pretrain}
\end{table}

\begin{table}[hbt!]
\caption{Model performance under different evaluation splits, reported as mean $\pm$ standard deviation. Metrics include mean relative absolute error (MRAE), mean absolute error (MAE), and coefficient of determination ($R^2$). Results are shown for models with and without computational pretraining where applicable. This is a supplement of Fig. 3c-f.} 
\centering
\small
\begin{tabular}{llccc}
\toprule
Split & Model & MRAE & MAE & R$^2$ \\
\midrule
Random split with pretraining& CGCNN & $\text{0.295}_\text{(0.014)}$ & $\text{0.603}_\text{(0.038)}$ & $\text{0.235}_\text{(0.068)}$ \\
 & CartNet & $\text{0.306}_\text{(0.009)}$ & $\text{0.606}_\text{(0.042)}$ & $\text{0.273}_\text{(0.062)}$ \\
 & ALIGNN & $\text{0.280}_\text{(0.013)}$ & $\text{0.550}_\text{(0.030)}$ & $\text{0.407}_\text{(0.051)}$ \\
 & CHGNet & $\textbf{\text{0.242}}_\text{(0.004)}$ & $\underline{\text{0.486}}_\text{(0.005)}$ & $\underline{\text{0.487}}_\text{(0.014)}$ \\
 & LEFTNet & $\underline{\text{0.242}}_\text{(0.003)}$ & $\textbf{\text{0.485}}_\text{(0.007)}$ & $\textbf{\text{0.507}}_\text{(0.017)}$ \\
\midrule
Random split without pretraining & CGCNN  & $\underline{\text{0.271}}_\text{(0.017)}$ & $\text{0.576}_\text{(0.019)}$ & $\underline{\text{0.313}}_\text{(0.051)}$ \\
 & CartNet & $\text{0.337}_\text{(0.015)}$ & $\text{0.738}_\text{(0.048)}$ & $\text{-0.342}_\text{(1.030)}$ \\
 & ALIGNN & $\text{0.286}_\text{(0.025)}$ & $\text{0.586}_\text{(0.054)}$ & $\text{0.369}_\text{(0.090)}$ \\
 & CHGNet & $\text{0.299}_\text{(0.010)}$ & $\text{0.613}_\text{(0.022)}$ & $\text{0.298}_\text{(0.056)}$ \\
 & LEFTNet & $\text{0.307}_\text{(0.012)}$ & $\text{0.616}_\text{(0.020)}$ & $\text{0.246}_\text{(0.130)}$ \\
 & LR & $\text{0.332}_\text{(0.005)}$ & $\text{0.637}_\text{(0.014)}$ & $\text{0.306}_\text{(0.008)}$ \\
 & RFR & $\text{0.284}_\text{(0.005)}$ & $\underline{\text{0.535}}_\text{(0.007)}$ & $\text{0.453}_\text{(0.016)}$ \\
 & SVR & $\textbf{\text{0.261}}_\text{(0.007)}$ & $\textbf{\text{0.510}}_\text{(0.013)}$ & $\textbf{\text{0.497}}_\text{(0.018)}$ \\
\midrule
Chemical system with pretraining& CGCNN & $\textbf{\text{0.275}}_\text{(0.029)}$ & $\text{0.562}_\text{(0.044)}$ & $\text{0.411}_\text{(0.148)}$ \\
 & CartNet & $\text{0.325}_\text{(0.051)}$ & $\text{0.617}_\text{(0.076)}$ & $\text{0.245}_\text{(0.285)}$ \\
 & ALIGNN & $\text{0.287}_\text{(0.033)}$ & $\text{0.576}_\text{(0.063)}$ & $\text{0.412}_\text{(0.118)}$ \\
 & CHGNet & $\text{0.287}_\text{(0.041)}$ & $\underline{\text{0.531}}_\text{(0.054)}$ & $\underline{\text{0.457}}_\text{(0.146)}$ \\
 & LEFTNet & $\underline{\text{0.283}}_\text{(0.042)}$ & $\textbf{\text{0.529}}_\text{(0.064)}$ & $\textbf{\text{0.484}}_\text{(0.135)}$ \\
\midrule
Chemical system without pretraining & CGCNN & $\text{0.302}_\text{(0.040)}$ & $\text{0.601}_\text{(0.074)}$ & $\text{0.345}_\text{(0.137)}$ \\
 & CartNet & $\text{0.328}_\text{(0.034)}$ & $\text{0.687}_\text{(0.069)}$ & $\text{0.183}_\text{(0.271)}$ \\
 & ALIGNN & $\underline{\text{0.295}}_\text{(0.045)}$ & $\text{0.594}_\text{(0.080)}$ & $\text{0.398}_\text{(0.125)}$ \\
 & CHGNet & $\text{0.366}_\text{(0.037)}$ & $\text{0.711}_\text{(0.135)}$ & $\text{0.197}_\text{(0.306)}$ \\
 & LEFTNet & $\text{0.331}_\text{(0.036)}$ & $\text{0.614}_\text{(0.037)}$ & $\text{0.372}_\text{(0.080)}$ \\
 & LR & $\text{0.361}_\text{(0.033)}$ & $\text{0.660}_\text{(0.047)}$ & $\text{0.328}_\text{(0.094)}$ \\
 & RFR & $\text{0.303}_\text{(0.044)}$ & $\underline{\text{0.550}}_\text{(0.061)}$ & $\textbf{\text{0.494}}_\text{(0.105)}$ \\
 & SVR & $\textbf{\text{0.294}}_\text{(0.046)}$ & $\textbf{\text{0.543}}_\text{(0.059)}$ & $\underline{\text{0.477}}_\text{(0.104)}$ \\
\midrule
Periodic groups with pretraining & CGCNN & $\text{0.326}_\text{(0.074)}$ & $\text{0.620}_\text{(0.096)}$ & $\text{0.287}_\text{(0.192)}$ \\
 & CartNet & $\text{0.345}_\text{(0.054)}$ & $\text{0.669}_\text{(0.100)}$ & $\text{-0.169}_\text{(0.795)}$ \\
 & ALIGNN & $\text{0.334}_\text{(0.061)}$ & $\text{0.640}_\text{(0.100)}$ & $\text{0.263}_\text{(0.193)}$ \\
 & CHGNet & $\textbf{\text{0.308}}_\text{(0.058)}$ & $\textbf{\text{0.566}}_\text{(0.083)}$ & $\underline{\text{0.346}}_\text{(0.258)}$ \\
 & LEFTNet & $\underline{\text{0.324}}_\text{(0.097)}$ & $\underline{\text{0.579}}_\text{(0.089)}$ & $\textbf{\text{0.377}}_\text{(0.174)}$ \\
\midrule
Periodic groups without pretraining & CGCNN & $\text{0.378}_\text{(0.077)}$ & $\underline{\text{0.716}}_\text{(0.083)}$ & $\text{0.096}_\text{(0.164)}$ \\
 & CartNet & $\text{0.382}_\text{(0.083)}$ & $\text{0.757}_\text{(0.092)}$ & $\text{-0.035}_\text{(0.289)}$ \\
 & ALIGNN & $\textbf{\text{0.342}}_\text{(0.052)}$ & $\textbf{\text{0.679}}_\text{(0.103)}$ & $\underline{\text{0.179}}_\text{(0.251)}$ \\
 & CHGNet & $\underline{\text{0.370}}_\text{(0.052)}$ & $\text{0.764}_\text{(0.131)}$ & $\text{-0.057}_\text{(0.387)}$ \\
 & LEFTNet & $\text{0.448}_\text{(0.183)}$ & $\text{0.832}_\text{(0.191)}$ & $\text{-0.223}_\text{(0.490)}$ \\
 & LR & $\text{0.481}_\text{(0.113)}$ & $\text{0.913}_\text{(0.261)}$ & $\text{-0.582}_\text{(0.885)}$ \\
 & RFR & $\text{0.386}_\text{(0.103)}$ & $\text{0.683}_\text{(0.096)}$ & $\textbf{\text{0.228}}_\text{(0.173)}$ \\
 & SVR & $\text{0.400}_\text{(0.097)}$ & $\text{0.773}_\text{(0.265)}$ & $\text{-0.051}_\text{(0.628)}$ \\
\midrule
Crystal system with pretraining & CGCNN & $\text{0.319}_\text{(0.084)}$ & $\text{0.619}_\text{(0.130)}$ & $\text{0.208}_\text{(0.402)}$ \\
 & CartNet & $\text{0.343}_\text{(0.071)}$ & $\text{0.672}_\text{(0.090)}$ & $\text{-0.104}_\text{(1.013)}$ \\
 & ALIGNN & $\text{0.298}_\text{(0.044)}$ & $\text{0.579}_\text{(0.086)}$ & $\text{0.431}_\text{(0.140)}$ \\
 & CHGNet & $\textbf{\text{0.282}}_\text{(0.046)}$ & $\textbf{\text{0.538}}_\text{(0.079)}$ & $\textbf{\text{0.483}}_\text{(0.128)}$ \\
 & LEFTNet & $\underline{\text{0.285}}_\text{(0.053)}$ & $\underline{\text{0.557}}_\text{(0.088)}$ & $\underline{\text{0.476}}_\text{(0.105)}$ \\
\midrule
Crystal system without pretraining & CGCNN & $\text{0.319}_\text{(0.046)}$ & $\text{0.617}_\text{(0.098)}$ & $\text{0.356}_\text{(0.163)}$ \\
 & CartNet & $\text{0.333}_\text{(0.065)}$ & $\text{0.746}_\text{(0.205)}$ & $\text{0.019}_\text{(0.482)}$ \\
 & ALIGNN & $\textbf{\text{0.304}}_\text{(0.039)}$ & $\text{0.643}_\text{(0.125)}$ & $\text{0.330}_\text{(0.213)}$ \\
 & CHGNet & $\text{0.375}_\text{(0.064)}$ & $\text{0.708}_\text{(0.144)}$ & $\text{0.237}_\text{(0.320)}$ \\
 & LEFTNet & $\text{0.344}_\text{(0.046)}$ & $\text{0.648}_\text{(0.081)}$ & $\text{0.338}_\text{(0.151)}$ \\
 & LR & $\text{0.379}_\text{(0.076)}$ & $\text{0.701}_\text{(0.130)}$ & $\text{0.236}_\text{(0.286)}$ \\
 & RFR & $\text{0.318}_\text{(0.049)}$ & $\underline{\text{0.587}}_\text{(0.087)}$ & $\textbf{\text{0.465}}_\text{(0.085)}$ \\
 & SVR & $\underline{\text{0.307}}_\text{(0.044)}$ & $\textbf{\text{0.575}}_\text{(0.080)}$ & $\underline{\text{0.444}}_\text{(0.123)}$ \\
\midrule
\bottomrule
\end{tabular}
\label{tab:materials_space_4panels}
\end{table}

\begin{table}[hbt!]
\centering
\caption{Pretrained model performance on the experimental test set (best in bold, second best underlined).}
\begin{tabular}{lccc}
\toprule
Model & MAE $\downarrow$ & MRAE $\downarrow$ & $R^2$ $\uparrow$ \\
\midrule
ALIGNN   & \underline{0.811} & \underline{0.346} &  -0.146 \\
CartNet    & 0.913 & 0.395 & \underline{-0.308} \\
CGCNN      & \textbf{0.627} & \textbf{0.295} & \textbf{0.271} \\
CHGNet   & 1.456 & 0.589 & -2.531 \\
LEFTNet  & 1.165 & 0.489 & -0.949 \\
\bottomrule
\end{tabular}
\label{tab:pretrained_models}
\end{table}

\begin{table}[hbtp]
\centering
\caption{Performance of different models on the validation and test sets, reported as mean $\pm$ standard deviation (\textbf{best in bold}, \underline{second best} in underline). Higher values indicate better performance ($\uparrow$), otherwise ($\downarrow$).}
\scalebox{1}{
\setlength{\tabcolsep}{1.0mm}{
\begin{tabular}{l|ccc|ccc}
\toprule
\multirow{2}{*}[-0.4ex]{Model} & \multicolumn{3}{c|}{Validation set} & \multicolumn{3}{c}{Test set} \\
& MAE(eV) $\downarrow$ & MRAE $\downarrow$ & ${R^2}$ $\uparrow$ & MAE(eV) $\downarrow$ & MRAE $\downarrow$ & ${R^2}$ $\uparrow$ \\
\midrule\midrule

\multicolumn{7}{l}{\textit{Without pretraining (fine-tuning only)}} \\ \midrule

CGCNN & $\text{0.566}_\text{(0.044)}$ & \textbf{$\text{0.277}_\text{(0.030)}$} & $\text{0.425}_\text{(0.092)}$ & $\text{0.576}_\text{(0.019)}$ & \underline{$\text{0.271}_\text{(0.017)}$} & $\text{0.313}_\text{(0.051)}$ \\
CartNet-Z & $\text{0.660}_\text{(0.092)}$ & $\text{0.322}_\text{(0.048)}$ & $\text{0.214}_\text{(0.221)}$ & $\text{0.738}_\text{(0.048)}$ & $\text{0.337}_\text{(0.015)}$ & $\text{-0.342}_\text{(1.030)}$ \\
CartNet-Prop & $\text{0.655}_\text{(0.109)}$ & $\text{0.315}_\text{(0.057)}$ & $\text{0.054}_\text{(0.656)}$ & $\text{0.703}_\text{(0.067)}$ & $\text{0.312}_\text{(0.023)}$ & $\text{-0.056}_\text{(0.434)}$ \\
ALIGNN-Z & $\text{0.598}_\text{(0.056)}$ & $\text{0.299}_\text{(0.036)}$ & $\text{0.401}_\text{(0.091)}$ & $\text{0.589}_\text{(0.041)}$ & $\text{0.279}_\text{(0.017)}$ & $\text{0.368}_\text{(0.073)}$ \\
ALIGNN-Prop & $\text{0.573}_\text{(0.073)}$ & \underline{$\text{0.288}_\text{(0.051)}$} & $\text{0.421}_\text{(0.117)}$ & $\text{0.586}_\text{(0.054)}$ & $\text{0.286}_\text{(0.025)}$ & $\text{0.369}_\text{(0.090)}$  \\
CHGNET-Z & $\text{0.634}_\text{(0.055)}$ & $\text{0.333}_\text{(0.038)}$ & $\text{0.300}_\text{(0.135)}$ & $\text{0.613}_\text{(0.022)}$ & $\text{0.299}_\text{(0.010)}$ & $\text{0.298}_\text{(0.056)}$ \\
CHGNET-Prop & $\text{0.635}_\text{(0.061)}$ & $\text{0.319}_\text{(0.040)}$ & $\text{0.325}_\text{(0.081)}$ & $\text{0.651}_\text{(0.042)}$ & $\text{0.316}_\text{(0.012)}$ & $\text{0.214}_\text{(0.091)}$ \\
LEFTNet-Z & $\text{0.614}_\text{(0.051)}$ & $\text{0.322}_\text{(0.043)}$ & $\text{0.361}_\text{(0.103)}$ & $\text{0.616}_\text{(0.020)}$ & $\text{0.307}_\text{(0.012)}$ & $\text{0.246}_\text{(0.130)}$ \\
LEFTNet-Prop & $\text{0.600}_\text{(0.104)}$ & $\text{0.304}_\text{(0.044)}$ & $\text{0.380}_\text{(0.183)}$ & $\text{0.612}_\text{(0.126)}$ & $\text{0.297}_\text{(0.029)}$ & $\text{0.256}_\text{(0.302)}$ \\

LR & $\text{0.650}_\text{(0.055)}$ & $\text{0.356}_\text{(0.049)}$ & $\text{0.357}_\text{(0.067)}$ & $\text{0.637}_\text{(0.014)}$ & $\text{0.332}_\text{(0.005)}$ & $\text{0.306}_\text{(0.008)}$ \\
RFR & \textbf{$\text{0.541}_\text{(0.047)}$} & $\text{0.296}_\text{(0.037)}$ & \textbf{$\text{0.510}_\text{(0.054)}$} & \underline{$\text{0.535}_\text{(0.007)}$} & $\text{0.284}_\text{(0.005)}$ & \underline{$\text{0.453}_\text{(0.016)}$} \\
SVR & \underline{$\text{0.544}_\text{(0.065)}$} & $\text{0.293}_\text{(0.051)}$ & \underline{$\text{0.480}_\text{(0.098)}$} & \textbf{$\text{0.510}_\text{(0.013)}$} & \textbf{$\text{0.261}_\text{(0.007)}$} & \textbf{$\text{0.497}_\text{(0.018)}$} \\ \midrule

\multicolumn{7}{l}{\textit{With pretraining}} \\ \midrule

CGCNN & $\text{0.621}_\text{(0.070)}$ & $\text{0.322}_\text{(0.050)}$ & $\text{0.227}_\text{(0.286)}$ & $\text{0.603}_\text{(0.038)}$ & $\text{0.295}_\text{(0.014)}$ & $\text{0.235}_\text{(0.068)}$ \\

CartNet-Z & $\text{0.613}_\text{(0.057)}$ & $\text{0.323}_\text{(0.033)}$ & $\text{0.264}_\text{(0.234)}$ & $\text{0.606}_\text{(0.042)}$ & $\text{0.306}_\text{(0.009)}$ & $\text{0.273}_\text{(0.062)}$ \\

CartNet-Prop & $\text{0.607}_\text{(0.079)}$ & $\text{0.314}_\text{(0.040)}$ & $\text{0.154}_\text{(0.506)}$ & $\text{0.601}_\text{(0.019)}$ & $\text{0.289}_\text{(0.007)}$ & $\text{0.147}_\text{(0.222)}$ \\
ALIGNN-Z & $\text{0.557}_\text{(0.028)}$ & $\text{0.278}_\text{(0.027)}$ & $\text{0.459}_\text{(0.059)}$ & $\text{0.581}_\text{(0.031)}$ & $\text{0.263}_\text{(0.013)}$ & $\text{0.338}_\text{(0.053)}$ \\

ALIGNN-Prop & $\text{0.529}_\text{(0.047)}$ & \textbf{$\text{0.266}_\text{(0.032)}$} & $\text{0.488}_\text{(0.096)}$ & $\text{0.550}_\text{(0.030)}$ & $\text{0.280}_\text{(0.013)}$ & $\text{0.407}_\text{(0.051)}$ \\

CHGNET-Z & $\text{0.535}_\text{(0.053)}$ & $\text{0.289}_\text{(0.039)}$ & $\text{0.472}_\text{(0.105)}$ & \underline{$\text{0.486}_\text{(0.005)}$} & \underline{$\text{0.242}_\text{(0.004)}$} & \underline{$\text{0.487}_\text{(0.014)}$} \\

CHGNET-Prop & \textbf{$\text{0.516}_\text{(0.038)}$} & $\text{0.276}_\text{(0.033)}$ & \underline{$\text{0.506}_\text{(0.091)}$} & $\text{0.525}_\text{(0.006)}$ & $\text{0.286}_\text{(0.004)}$ & $\text{0.406}_\text{(0.014)}$ \\

LEFTNet-Z & \underline{$\text{0.520}_\text{(0.044)}$} & $\text{0.280}_\text{(0.040)}$ & \textbf{$\text{0.512}_\text{(0.101)}$} & \textbf{$\text{0.485}_\text{(0.007)}$} & $\text{0.242}_\text{(0.003)}$ & \textbf{$\text{0.507}_\text{(0.017)}$} \\

LEFTNet-Prop & $\text{0.528}_\text{(0.055)}$ & \underline{$\text{0.268}_\text{(0.044)}$} & $\text{0.495}_\text{(0.089)}$ & $\text{0.494}_\text{(0.040)}$ & \textbf{$\text{0.232}_\text{(0.013)}$} & $\text{0.484}_\text{(0.061)}$ \\

\midrule\midrule

\bottomrule
\end{tabular}}}
\label{tab:cv_performance}
\vspace{-15pt}
\end{table}

\begin{sidewaystable}[htbp]
\caption{LOMO evaluation comparing Z encoding and atomic property embeddings with and without pretraining (MRAE).}
\centering
\setlength{\tabcolsep}{4pt}
\renewcommand{\arraystretch}{1.15}

{\bfseries With pretraining}

\vspace{3mm}

\begin{tabular}{llcccccccccc}
\toprule
Model & Encoding & Antimonides & Arsenides & Carbides & Chalcogenides &
Halides & Hydrides & Nitrides & Oxides & Phosphides & Silicides \\
\midrule
CartNet & Z & 0.716 & 0.401 & 0.250 & 0.318 & 0.422 & 0.542 & 0.824 & 0.371 & 0.435 & 0.619 \\
CartNet & Prop & 0.267 & 0.414 & 0.322 & 0.297 & 0.423 & 0.508 & 0.520 & 0.497 & 0.319 & 1.008 \\
\midrule
ALIGNN & Z & 0.448 & 0.358 & 0.190 & 0.312 & 0.367 & 0.307 & 0.496 & 0.332 & 0.324 & 0.664 \\
ALIGNN & Prop & 0.248 & 0.381 & 0.272 & 0.312 & 0.402 & 0.394 & 0.509 & 0.340 & 0.394 & 0.383 \\
\midrule
LEFTNet & Z & 0.555 & 0.498 & 0.255 & 0.269 & 0.363 & 0.299 & 0.488 & 0.286 & 0.369 & 0.896 \\
LEFTNet & Prop & 0.309 & 0.348 & 0.271 & 0.275 & 0.334 & 0.222 & 0.520 & 0.294 & 0.334 & 0.346 \\
\midrule
CHGNet & Z & 0.443 & 0.470 & 0.236 & 0.279 & 0.410 & 0.355 & 0.444 & 0.287 & 0.369 & 0.972 \\
CHGNet & Prop & 0.333 & 0.387 & 0.266 & 0.264 & 0.414 & 0.297 & 0.591 & 0.289 & 0.289 & 0.692 \\
\bottomrule
\end{tabular}

\vspace{8mm}

{\bfseries Without pretraining}

\vspace{3mm}

\begin{tabular}{llcccccccccc}
\toprule
Model & Encoding & Antimonides & Arsenides & Carbides & Chalcogenides &
Halides & Hydrides & Nitrides & Oxides & Phosphides & Silicides \\
\midrule
CartNet & Z & 0.387 & 0.358 & 0.106 & 0.365 & 0.370 & 0.379 & 0.467 & 0.351 & 0.323 & 1.072 \\
CartNet & Prop & 0.435 & 0.498 & 0.121 & 0.359 & 0.387 & 0.488 & 0.421 & 0.361 & 0.244 & 0.907 \\
\midrule
ALIGNN & Z & 0.350 & 0.289 & 0.207 & 0.342 & 0.449 & 0.392 & 0.408 & 0.314 & 0.376 & 0.400 \\
ALIGNN & Prop & 0.350 & 0.289 & 0.207 & 0.342 & 0.449 & 0.392 & 0.408 & 0.314 & 0.376 & 0.400 \\
\midrule
LEFTNet & Z & 0.687 & 0.620 & 0.199 & 0.447 & 0.441 & 0.395 & 0.465 & 0.334 & 0.291 & 0.457 \\
LEFTNet & Prop & 0.500 & 0.438 & 0.199 & 0.327 & 0.507 & 0.364 & 0.453 & 0.325 & 0.333 & 0.540 \\
\midrule
CHGNet & Z & 0.376 & 0.426 & 0.157 & 0.372 & 0.385 & 0.427 & 0.468 & 0.344 & 0.388 & 0.360 \\
CHGNet & Prop & 0.349 & 0.384 & 0.159 & 0.378 & 0.407 & 0.376 & 0.487 & 0.346 & 0.272 & 0.306 \\
\bottomrule
\end{tabular}

\label{tab:encoding_lomo}
\end{sidewaystable}

\FloatBarrier
\section{Classical machine learning baselines on learned representations}
\begin{figure}
    \centering
    \includegraphics[width=\linewidth]{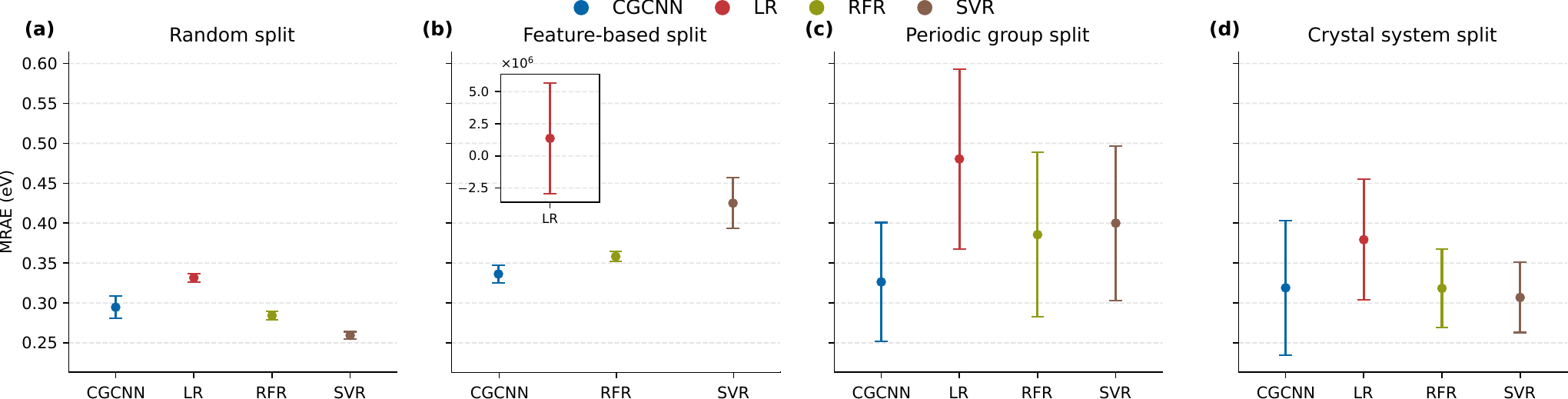}
    \caption{Test MRAE of CGCNN and classical machine learning baselines (LR, RFR, SVR) under different data splits: \textbf{a,} random split, \textbf{b,} feature-based OOD split (LR shown in the inset due to the difference in scale), \textbf{c,} periodic group split, and \textbf{d,} crystal system split. Error bars indicate one standard deviation across runs.}
    \label{fig:classical_ml_compare}
\end{figure}
We considered linear regression (LR), random forest regression (RFR) and support vector regression (SVR), representing linear, kernel-based, and tree-based learning paradigms, respectively. These models are used as probes to assess how much predictive signal is already encoded in the learned representation. We evaluated the models under the random split (Supplementary Figure~\ref{fig:classical_ml_compare}a), feature-based split (Supplementary Figure~\ref{fig:classical_ml_compare}b), periodic group split as a representative chemistry-based split (Supplementary Figure~\ref{fig:classical_ml_compare}c), and crystal-system split as a structure-based split (Supplementary Figure~\ref{fig:classical_ml_compare}d).

Across all evaluated tasks, we observed that nonlinear modeling (SVR and RFR) consistently outperformed LR. This indicates that linear mappings are insufficient to exploit the predictive information contained in the embedding fully.  Under feature-based OOD, nonlinear models maintain strong performance, while the LR fails the task. This suggests that the predictive information in the embeddings retains local consistency in the held-out set, reflecting the role of local chemical and coordination motifs in determining electronic structure.

Among the nonlinear models, their relative strengths depend on the difficulty of the task. Under relatively easy settings, such as random and crystal-system splits, SVR achieves slightly better performance, likely benefiting from its smooth kernel-based interpolation. In contrast, under more challenging OOD scenarios, particularly feature-based and periodic-group splits, RFR exhibits stronger robustness, suggesting that segmentation-based models generalize more reliably when test samples exceed the training set.

At the same time, the performance of classical models, sometimes even rivaling or surpassing certain GNNs, should be interpreted with caution. These results do not directly indicate whether such models can capture material physics. Because the encoding is obtained at the compound level by averaging the atomic environment, the model acquires not an explicit crystal geometry, but rather an aggregated representation of structural information. Instead, these results indicate that the predictive information encoded in the learned representation is strongly correlated with the bandgap and can be effectively exploited for estimation by relatively simple models. 

\section{Experimental results on atomic encoding SHAP values of SVR}

This section supplements Fig. 5 by providing the complete SHAP summary tables for the SVR model across all encoding features. The features are grouped into three segments: atom-level self features, neighbor-aggregated features, and radial basis function (RBF) distance features. For each feature, we report the mean SHAP value and the empirical distribution of the corresponding feature values in the training set. Feature distributions are normalized at the material level, where 0 indicates that none of the atoms in a given material falls into the corresponding bin and 1 indicates that all atoms fall into that bin.
\clearpage

\includepdf[
  pages=1-5,
  scale=0.73,
  pagecommand={
    \noindent
    \makebox[\textwidth][r]{%
      \raisebox{-\dimexpr\textheight-\footskip+1cm\relax}[0pt][0pt]{%
        \footnotesize(figure continued on next page)%
      }%
    }
  }
]{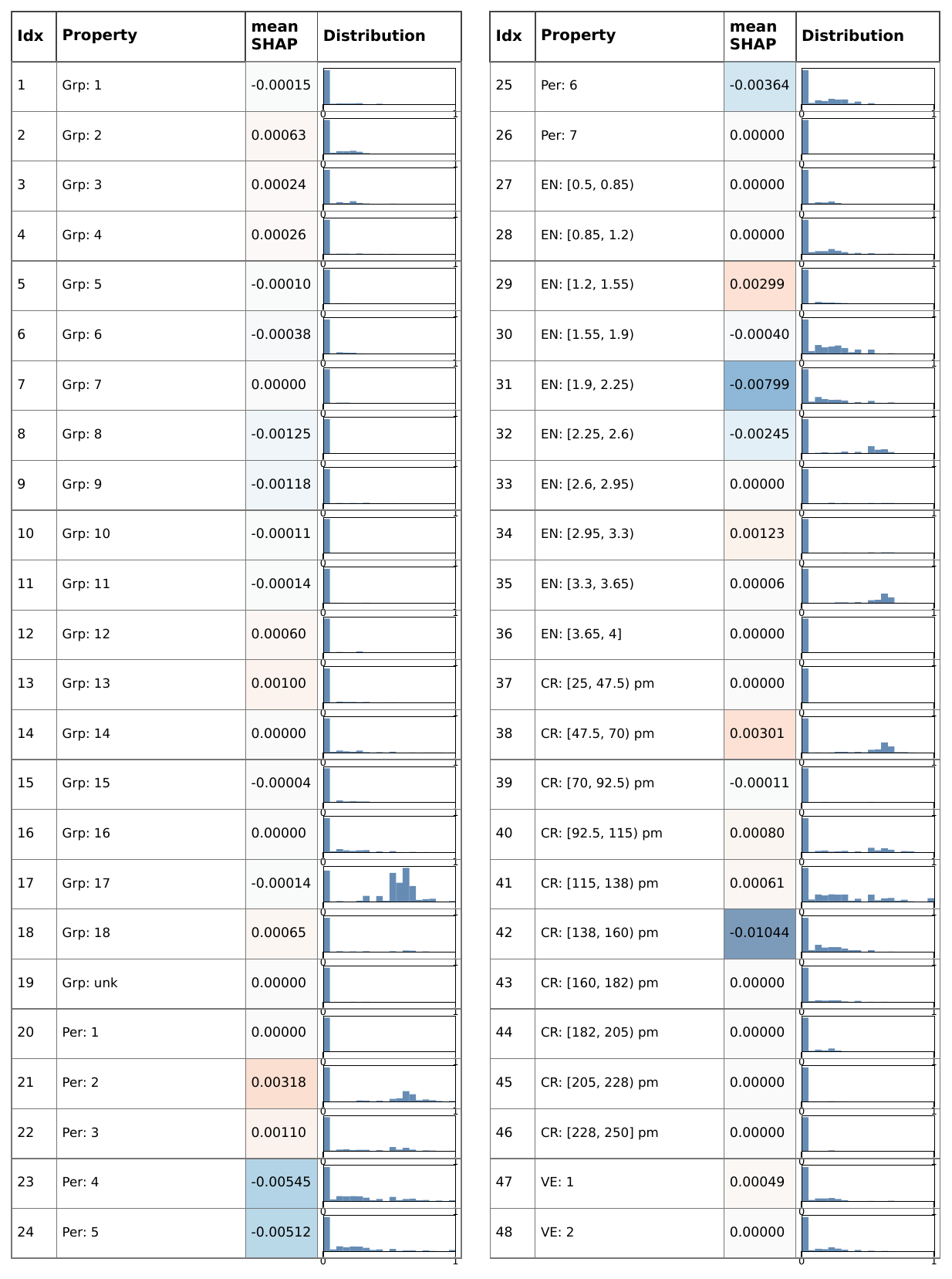}

\includepdf[
  pages=6,
  scale=0.73,
  pagecommand={
   \captionsetup{type=figure}
    \vspace{17cm}
    \caption{Average SHAP value and distribution for SVR across all segments. The feature abbreviations are defined as follows: Grp (group number), Per (period number), EN (electronegativity), CR (covalent radius), VE (valence electrons), FIE (first ionization energy), EA (electron affinity), Blk (block), and AV (atomic volume). Features are grouped into three segments: self-atom features (Idx 1--92), neighbor-atom features (Idx 93--184), and encoded distance features (Idx 185--220). In each row, the Distribution panel shows the distribution of training-set values for the corresponding feature on the normalized 0--1 input scale; taller bars indicate that more training samples fall within that interval.}
    \label{fig:shap_svr}
  }
]{figures/supplementary/shap_feature_table_svr_signed_all_segments.pdf}

\ifdefined\mergedintomain
\else
\end{document}
\fi
\end{document}